\documentclass[aps,preprint,superscriptaddress,nofootinbib]{revtex4}%
\usepackage{amsfonts}
\usepackage{amsmath}
\usepackage{amssymb}
\usepackage{graphicx}%
\newcommand{\be}{\begin{equation}}
\newcommand{\ee}{\end{equation}}
\newcommand{\bea}{\begin{eqnarray}}
\newcommand{\eea}{\end{eqnarray}}

\begin{document}

\title[The $\Sigma\Phi EA$ model]
    {\textbf{SCALAR AND TENSORIAL TOPOLOGICAL MATTER COUPLED TO (2+1)-DIMENSIONAL GRAVITY:\\
            A. CLASSICAL THEORY AND GLOBAL CHARGES}}

\author{\textbf{R.\ B.\ Mann}}

    \email{rbmann@sciborg.uwaterloo.ca}

    \affiliation{Perimeter Institute for Theoretical Physics, Waterloo,
                 Ontario, Canada, N2J 2G9}

    \affiliation{Department of Physics, University of Waterloo, Waterloo,
                 Ontario, Canada N2L 3G1}

\author{\textbf{Eugeniu M.\ Popescu}}

    \email{empopesc@uwaterloo.ca}

    \affiliation{Department of Physics, University of Waterloo, Waterloo,
                 Ontario, Canada N2L 3G1}


\begin{abstract}
    \noindent
        We consider the coupling of scalar topological matter to (2+1)-dimensional
        gravity. The matter fields consist of  a 0-form scalar field and a
        2-form tensor field. We carry out a canonical analysis of the classical theory,
        investigating its sectors and solutions. We show that the model admits both
        BTZ-like black-hole solutions and homogeneous/inhomogeneous FRW cosmological
        solutions.We also investigate the global charges associated with the model
        and show that the algebra of charges is the extension of the Ka\v{c}-Moody
        algebra for the field-rigid gauge charges, and the Virasoro algebra for the
        diffeomorphism charges. Finally, we show that the model can be written as a
        generalized Chern-Simons theory, opening the perspective for its formulation as a
        generalized higher gauge theory.
\end{abstract}


\date[Date: ]{November $27^{th}$, 2005}

\startpage{1}
\endpage{25}

\maketitle


\section{Introduction}

\indent
    Despite its apparent simplicity, and despite the fact that
    it has been shown to to be exactly solvable in the absence
    of matter \cite{Witten1}, coupling matter to (2+1)-dimensional
    gravity in the traditional way generally destroys its
    solvability properties. In this case the quantization process
    is once again faced with much the same issues as its
    (3+1)-dimensional counterpart.\\
\indent
    In the attempts to study the quantization of (2+1)-dimensional
    gravity in the presence of matter, the BCEA model has emerged
    as one of the few theories in which matter can be coupled to
    gravity while still preserving the solvability inherited from
    the latter. Proposed originally as a soluble diffeomorphism
    invariant theory \cite{Horowitz} and later studied in a slightly
    modified form in \cite{bcea}, the BCEA model is essentially a
    topological field theory in which 1-form matter fields are
    minimally coupled to gravity in the first-order formalism through
    the connection. Coupling matter to gravity in this
    non-traditional\footnote[1]{We refer to the coupling of matter
    through the connection as `non-traditional' as opposed to the
    `traditional' coupling of matter to gravity through the metric in the
    metric formalism, or alternatively, through the triad and/or
    co-triad fields in the first-order formalism.} way has the
    effect of introducing only a finite number of degrees of
    freedom in addition to those of pure gravity, such that the
    resulting phase-space of the theory remains
    finite-dimensional and hence solvable both classically and
    quantum mechanically.\\
\indent
    What makes this model interesting from the physical viewpoint
    is the fact that for nontrivial topologies, it has non-trivial
    solutions. In particular, it has been shown  that it admits as
    a solution the BTZ black-hole geometry \cite{bceabtz} with the
    surprising result that the Noether charges in this case - the
    quasilocal energy and angular momentum - change roles as
    compared to their counterparts in Einsteinian gravity.
    Explicitly, the quasilocal energy in the BTZ theory is
    proportional to the quasilocal angular momentum parameter in
    Einsteinian gravity with negative cosmological constant, and
    vice-versa. Furthermore, it has been shown in \cite{bcea} that
    the model can be written as a Chern-Simons theory with the
    group $I[ISO(2,1)]$ obtained directly from the Lie algebra of
    the constraints. Notwithstanding computational difficulties,
    this makes the model quantizable in a rather straightforward
    manner for any topology of relevance, and in particular for
    the topology of the BTZ black-hole solution mentioned
    earlier.\\
\indent
    The BCEA theory, however, is not the only theory in which
    matter can be coupled minimally to gravitation in (2+1)
    dimensions. Another obvious possibility is to construct a model
    where instead of coupling two 1-form matter fields to gravity
    as in the BCEA case, we couple to gravity a 2-form field and
    a 0-form field.\\
\indent
    In the present paper we consider the latter kind of theory,
    which will subsequently be called the $\Sigma\Phi EA$ theory,
    and we compare its results to the corresponding results
    in the BCEA model. Of course, one expects the two theories to
    have strong similarities, and in the following we will show that
    indeed this is the case. Both theories admit the BTZ geometry as
    a solution, and both theories exhibit similar anomalies regarding
    the Noether charges. Nevertheless, there are also major
    differences between the two models. Aside from the fact that
    they have different constraint algebras, while the BCEA model
    can be written in a straightforward manner as a Chern-Simons
    theory with the group generated by its constraint algebra,
    for the new model the situation is more complicated
    and in fact, more interesting. In  the latter case, the constraint
    algebra becomes insufficient for writing the
    theory as a Chern-Simons theory, and one has to make
    additional use of the quaternion algebra for this purpose.
    Needless to say, this approach of writing the $\Sigma\Phi EA$
    model as a Chern-Simons theory raises very interesting
    questions regarding the canonical analysis and the quantization
    of the model.\\
\indent
    The paper is organized as follows. In Section II, we review
    briefly the relevant results of the BCEA theory, for the purpose
    of later comparison with corresponding results of the $\Sigma\Phi EA$
    model, which is described in detail in Section III. In Section IV
    we discuss the classical sectors of the theory, and in Section V we
    discuss certain solutions of the theory that illustrate these sectors.
    In Section VI we discuss in detail the global gauge charges of
    the theory, and we determine the classical and operator
    version of the algebras of gauge and diffeomorphism charges.
    In Section VII we show that the the $\Sigma\Phi EA$ theory can be
    written as a generalized Chern-Simons theory in the manner described
    above, and in Section VIII we conclude with a discussion of
    the results and issues that emerge from our analysis.

\bigskip

\section{Review of the BCEA model}

\indent
    In this section, we briefly review the relevant results of the BCEA
    theory as they pertain to the purpose of this paper. For more
    details, the reader should consult \cite{bcea},
    \cite{bceabtz}.\\
\indent
    The action of the BCEA model in the first order formulation
    has the expression:
        \be
            S[B,C,E,A]=\int_{M}(E_{i}\wedge R^{i}[A]+B_{i}\wedge DC^{i})
            \label{sbcea}
        \ee
    where $M$ is a 3-dimensional non-compact spacetime
    with the the topology $M$=$R\times S$, and $S$
    is a 2-dimensional spacelike surface. The fields $E_{i}$
    in (\ref{sbcea}) are $SO(2,1)$ 1-forms which, if invertible,
    correspond to the triads of the spacetime metric, and $R^{i}[A]$
    are the curvature 2-forms associated to the $SO(2,1)$ connection
    1-forms $A^{i}$, with the expression:
        \be
            R^{i}[A]=dA^{i}+\frac{1}{2}\,\epsilon^{ijk}A_{j}\wedge A^{k}
            \label{RA}
        \ee
    The $SO(2,1)$ 1-forms $B^{i}$, $C^{i}$, are the topological
    matter fields that are coupled to the fields $E^{i}$, $A^{i}$
    of pure gravity, and $DC^{i}$ is the covariant derivative of
    the field $C^{i}$, having the expression:
        \be
            DC^{i}=dC^{i}+\epsilon^{ijk}A_{j}\wedge C_{k}\label{covder}
        \ee
    Throughout the entire paper we adopt the following index
    convention. Greek indices, taking the values $0$, $1$, $2$,
    designate the spacetime components of tensors, and are raised
    and lowered by the spacetime metric $g_{\alpha,\beta}$. Latin
    lower case indices, also taking the values $0$, $1$, $2$, are
    $SO(2,1)$ indices,and are raised and lowered by the $SO(2,1)$
    metric $\eta_{ij}$=$diag(-1,1,1)$, and $\epsilon^{ijk}$ is
    the totally antisymmetric $SO(2,1)$ symbol with $\epsilon^{012}$=$1$.
    Any other type of indices that might appear in the paper will be
    appropriately explained in the context where they  occur.\\
\indent
    The action (\ref{sbcea}) yields, upon first order variation
    (and up to surface terms), the equations of motion:
        \bea
            R^{i}[A]&=&0\nonumber\\
            DE^{i}+\epsilon^{ijk}B_{j}\wedge C_{k}&=&0\\
            DB^{i}=DC^{i}&=&0\nonumber
            \label{eqnmotn}
        \eea
    and is invariant under the following 12-parameter
    infinitesimal gauge transformations:
        \bea
            \delta A^{i}&=&D\tau^{i}\nonumber\\
            \delta B^{i}&=&D\rho^{i}+\epsilon^{ijk}B_{j}\tau_{k}\nonumber\\
            \delta C^{i}&=&D\lambda^{i}+\epsilon^{ijk}C_{j}\tau_{k}\\
            \delta E^{i}&=&D\beta^{i}+\epsilon^{ijk}(E_{j}\tau_{k}+
                            B_{j}\lambda_{k}+C_{j}\rho_{k})\nonumber
            \label{eqngauge}
        \eea
    where $\beta^{i}$, $\lambda^{i}$, $\rho^{i}$, $\tau^{i}$ are
    0-form gauge parameters.\\
\indent
    The $(2+1)$ canonical splitting induced by the topology of the
    manifold $M$ yields four sets of constraints ${J^{i}}$, ${P^{i}}$,
    ${Q^{i}}$, ${R^{i}}$, which are enforced by the zeroth spacetime
    components of the form fields $A^{i}$, $E^{i}$, $B^{i}$, and $C^{i}$
    respectively, acting as Lagrange multipliers. The Lie algebra generated
    by these constraints is:
        \bea
            \{J^{i},J^{j}\}=\epsilon^{ijk} J_{k};\,
            \{J^{i},P^{j}\}=\epsilon^{ijk} P_{k};\,
            \{J^{i},Q^{j}\}=\epsilon^{ijk} Q_{k}\nonumber\\
            \{J^{i},R^{j}\}=\epsilon^{ijk} R_{k};\,
            \{Q^{i},R^{j}\}=\epsilon^{ijk} P_{k}\label{alg}
        \eea
    with the rest of the Poisson brackets being zero. Equations
    (\ref{alg}) can be recognized as the Lie algebra of the
    inhomogenized Poincar\'{e} group $I[ISO(2,1)]$ \cite{GKL}. The
    Hamiltonian of the system is zero on shell, since it
    depends only on the constraints, and consequently the
    constraints are preserved in time.\\
\indent
    As mentioned earlier, the BCEA theory admits the BTZ
    black-hole geometry as a solution. Taking into account the
    symplectic structure generated by the BCEA action
    (\ref{sbcea}), the conserved charges for the BTZ black-hole in
    this theory are found to be \cite{bceabtz}
        \bea
            \cal{M}_{BCEA}= \frac{\pi \cal{J}}{\textit{l}}\nonumber\\
            \cal{J}_{BCEA}= -\pi \cal{M}\textit{l}\label{bcbtzcharges}
        \eea
    where $\cal{M}$ and $\cal{J}$ are respectively conserved mass
    and the angular momentum of the BTZ black-hole in Einsteinian
    gravity with negative cosmological constant, and $\textit{l}$
    is related to the cosmological constant $\Lambda$ through the
    relation:
        \be
            \Lambda = -\frac{1}{\textit{l}^{2}}
        \ee
    Note that the cosmological constant is a constant of
    integration in this theory, and not a parameter in the
    action.

\bigskip

\section{The $\Sigma\Phi EA$ Model}

\indent
    In this section, we define the $\Sigma\Phi EA$ theory
    and analyze its classical properties, highlighting both
    the similarities and the differences between this theory
    and the BCEA model.

\subsection{Action, equations of motion and gauge symmetries}

\indent
    The action of the $\Sigma\Phi EA$ model is defined analogously to the BCEA model:
        \be
            S[\Sigma,\Phi,E,A]=\int_{M}(E_{i}\wedge R^{i}[A]+\Sigma_{i}\wedge D\Phi^{i})
            \label{ssfea}
        \ee
    where the fields $E^{i}$, $A^{i}$ and the covariant derivative
    have the same significance as in the BCEA theory, and the fields
    $\Sigma^{i}$, $\Phi^{i}$ are now respectively $SO(2,1)$-valued
    2-form and a 0-form matter fields coupled to gravity through
    the connection $A^{i}$ in the covariant derivative.\\
\indent
    Up to surface terms, the first order variation of the action
    (\ref{ssfea}) yields the equations of motion:
         \bea
            R^{i}[A]&=&0\nonumber\\
            DE^{i}+\epsilon^{ijk}\Sigma_{j}\wedge \Phi_{k}&=&0\\
            D\Sigma^{i}=D\Phi^{i}&=&0\nonumber \label{eqnmotn1}
        \eea
    which are, as expected, very similar to the equations of
    motion (\ref{eqnmotn}) of the BCEA theory.\\
\indent
    The equations of motion (\ref{eqnmotn1}) are invariant under
    the following infinitesimal gauge transformations:
        \bea
            \delta A^{i}&=&D\alpha^{i}\nonumber\\
            \delta \Phi^{i}&=&\epsilon^{ijk}\Phi_{j}\alpha_{k}\nonumber\\
            \delta \Sigma^{i}&=&D\gamma^{i}+\epsilon^{ijk}\Sigma_{j}\alpha_{k}\\
            \delta E^{i}&=&D\beta^{i}+\epsilon^{ijk}(E_{j}\alpha_{k}
                            -\Phi_{j}\gamma_{k})\nonumber
            \label{eqngauge1}
        \eea
    with $\alpha^{i}$, $\beta^{i}$ 0-form and $\gamma^{i}$
    1-form gauge parameters. It would appear from
    (\ref{eqngauge1}) that the equations of motion of the $\Sigma\Phi
    EA$ theory are invariant under a 15-parameter set of gauge
    transformations. This is however not the case since the gauge
    transformations are themselves invariant under the
    infinitesimal ``translation":
        \be
            {\gamma^{'}}^{i}=\gamma^{i}+D\eta^{i}\label{cosmconst}
        \ee
    which reduces the number of independent gauge parameters to
    12.\\
\indent
    Splitting the action (\ref{ssfea}) in accordance with the
    topology of the manifold $M=R\times S$, yields the expression:
        \bea
            S[\Sigma,\Phi,E,A]=\int_{R}dt\int_{S}d^{2}x[\tilde{E}_{i}^{B}\dot{A}^{i}_{B}+
            \frac{1}{2}\tilde{\Sigma}_{i}\dot{\Phi}^{i}+A_{i0}J^{i}+E_{i0}P^{i}+
            \Sigma_{i0A}K^{iA}]\label{actsplit}
        \eea
    where Latin uppercase indices are spacelike indices taking
    the values $1,2$, tilded quantities are densitized fields with
    $\epsilon^{AB}=\epsilon^{0AB}$, and dotted quantities are are
    the time derivatives of the corresponding fields. As expected,
    the spatial components of the form-fields form pairs of
    canonically conjugate variables, and the zeroth components of
    the fields act as Lagrange multipliers enforcing the
    constraints:
        \bea
            J^{i}&=&\star(\hat{D}\hat{E}^{i}-\epsilon^{ijk}\hat{\Sigma}_{j}
                   \hat{\Phi}_{k})\nonumber\\
            P^{i}&=&\star(R^{i}[\hat{A}])\label{constraints}\\
            K^{iA}&=&[\star(\hat{D}\hat{\Phi}^{i})]^{A}\nonumber
        \eea
    where $\star$ is the spatial Hodge dual and the caret signifies the projection
    of the corresponding quantity onto the spacelike surface $S$. We see from
    (\ref{constraints}) that the model has 12 independent first-class constraints,
    consistent with the number of independent gauge parameters found earlier.\\
\indent
    Relabeling the constraints $K^{i1}$ and $K^{i2}$ by $Q^{i}$ and respectvely
    $R^{i}$ - the order of the relabeling will prove to be irrelevant - for easier
    comparison with the BCEA model, a tedious but straightforward calculation yields
    a Poisson constraint algebra almost identical to the constraint algebra
    (\ref{alg}). All the Poisson brackets of the constraints are identical to the
    corresponding brackets of the BCEA model except for the bracket of $Q^{i}$ with
    $R^{i}$ which is now given by the expression:
        \be
            \{Q^{i},R^{j}\}=0 \label{sfeaalg}
        \ee
    Consequently, the constraint algebra of the $\Sigma\Phi EA$
    model can be viewed as the Lie algebra of the
    (2+1)-dimensional Lorentz group $SO(2,1)$ generated by
    $\{J^{i}\}$ inhomogeneized by three sets of Poincar\'{e}
    translation-like abelian generators $\{P^{i}\}$, $\{Q^{i}\}$,
    $\{R^{i}\}$ that also commute with each other. In the absence
    of any nomenclature regarding the particular types of
    inhomogenization of simple groups, we have decided to use for
    the group associated with the constraint algebra of this model
    the obvious notation $[3I]SO(2,1)$, where $[1I]SO(2,1)\equiv ISO(2,1)$
    is the (2+1)-dimensional Poincar\'{e} group.\\
\indent
    One should note at this time the following interesting aspect of
    the constraints in the $\Sigma\Phi EA$ model. In the BCEA model
    both matter fields generate independent symmetries through the
    constraints $Q^{i}$, $R^{i}$. However in the $\Sigma\Phi EA$ model the
    symmetries associated with the matter fields are generated,
    surprisingly, only by the scalar fields $\Phi^{i}$, as is
    clearly shown by (\ref{constraints}). Both matter fields couple as expected
    to the symmetry generators $P^{i}$, but in the $\Sigma\Phi EA$
    theory there are no symmetries generated exclusively by the
    2-form matter fields $\Sigma^{i}$. This fact suggests that any
    nontrivial solution of the theory should have nontrivial scalar
    fields, at least globally if not locally. This is a specific
    characteristic of the $\Sigma\Phi EA$ model; no such argument regarding
    nontrivial solutions can be made for BCEA theory.

\subsection{Degrees of freedom}

\indent
    At this time, a natural question to ask is whether the $\Sigma\Phi EA$
    theory has any local degrees of freedom (introduced by the topological
    matter fields) or not. Since the answer to this question involves a
    detailed analysis of the constraints and symmetries determined earlier,
    we will list them below in explicit form, for future reference. With the
    relabeling of the K-constraints introduced in the previous subsection,
    the explicit form of the constraints (\ref{constraints}) is given by the
    relations:
        \bea
            J^{i}&=&\partial_{[1}E^{i}_{2]}+\epsilon^{ijk}(A_{j[1}E_{k2]}-
                                                \Sigma_{j[12]}\Phi_{k})\nonumber\\
            P^{i}&=&\partial_{[1}A^{i}_{2]}+
                                \frac{1}{2}\,\epsilon^{ijk}A_{j[1}A_{k2]}\nonumber\\
                        Q^{i}&=&\partial_{1}\Phi^{i}+\epsilon^{ijk}A_{j1}\Phi_{k}\label{constraints1}\nonumber\\
            R^{i}&=&\partial_{2}\Phi^{i}+\epsilon^{ijk}A_{j2}\Phi_{k}
        \eea
    where the antisymmetrization involves only the arabic numeral
    indices designating the spatial components of the fields. The variables
    in (\ref{constraints1}) are invariant under the gauge transformations:
        \bea
            \delta A^{i}_{a}&=&\partial_{a}\tau^{i}+\epsilon^{ijk}A_{ja}\tau_{k}\nonumber\\
            \delta \Phi^{i}&=&\epsilon^{ijk}\Phi_{j}\tau_{k}\nonumber\\
            \delta \Sigma^{i}_{[ab]}&=&\partial_{[a}\lambda^{i}_{b]}+
                        \epsilon^{ijk}(A_{j[a}\lambda_{kb]}+\Sigma_{j[ab]}\tau_{k})\nonumber\label{eqngauge2}\\
            \delta E^{i}_{a}&=&\partial_{a}\beta^{i}+\epsilon^{ijk}(A_{ja}\beta_{k}+
                            E_{j}\tau_{k}-\Phi_{j}\lambda_{ka})
        \eea
    where once again, the antisymmetrization operation involves only the
    spatial indices of the fields $a,b=1,2$.\\
\indent
    Returning now to the issue of the physical degrees of freedom,
    it is well-known that the major ingredients in determining the
    number of physical degrees of freedom (PDOF) of a system are the total
    number of canonical variables (CV), the total number of independent first
    class (IFCC) and second class constraints (ISCC), and the total number of
    independent conditions one can impose on the system in order
    to fix the gauge (IGC). Once these ingredients are known, the
    number of physical degrees of freedom of the system is given
    by the relation \cite{Teit}:
        \be
            (\#PDOF)=(\#CV)-(\#IFCC)-(\#ISCC)-(\#IGC)\label{DOFeqn}
        \ee
    Therefore, in order to establish the number of physical degrees of freedom,
    and since the $\Sigma\Phi EA$ theory has no second class constraints\footnote{For
    the purpose of this analysis we restrict ourselves to the Lagrangian
    formalism where no second class constraints are present in this
    model}, we need to determine the number of independent first
    class constraints and the corresponding number of independent
    gauge fixing conditions.\\
\indent
    That such a step is necessary at this stage of the analysis becomes obvious
    if one attempts to determine the number of degrees of freedom based on the
    $prima facie$ information contained in the above constraints and gauge
    symmetries. The total number of canonical variables as determined from
    (\ref{actsplit}) is $(\#CV)=18$, and from (\ref{constraints1}) and
    (\ref{eqngauge2}), one would have twelve independent first class constraints
    $(\#IFCC)=12$ and similarly twelve independent gauge fixing conditions
    $(\#IGC)=12$. Under these circumstances, (\ref{DOFeqn}) would yield for
    the number of physical degrees of freedom of the $\Sigma\Phi EA$
    theory a negative number $(\#PDOF)=-6$. Such a result, while not
    impossible (for example in the case of (1+1)-dimensional
    gravity the number of physical degrees of freedom is also
    negative), is a very strong indication that the
    constraints and/or the gauge fixing conditions might not be
    all independent as assumed.\\
\indent
    In order to establish whether this is indeed the case we proceed
    to investigate the constraints and the gauge fixing conditions
    separately.
        \begin{enumerate}
            \item[{\textbf{(i).}}]{\textbf{The constraints}}\\
                The first thing that should be noted is the fact that the fields
                $\Phi^{i}$ which are scalar forms appear in two sets of the constraints
                (\ref{constraints1}), namely in $Q^{i}$ and $R^{i}$, together with the
                spatial components of the spin connection, suggesting that these two
                constraints together with the constraints $J^{i}$ might be connected.
                This is indeed the case, and using the cohomological properties of the
                exterior derivative it is not difficult to show that between these
                three sets of constraints one has the relation:
                    \be
                        \partial_{2}Q^{i}-\partial_{1}R^{i}+\epsilon^{ijk}(P_{j}\Phi_{k}+
                            A_{j2}Q_{k}-A_{j1}R_{k})=0\label{eqndepcons}
                    \ee
                where the equality in (\ref{eqndepcons}) is a strong equality, i.e. the
                above relation is also valid off-shell. Under these
                circumstances, and since the remaining constraints of the theory are
                independent, it follows that in reality there are only nine independent
                first class constraints instead of twelve, and consequently $(\#IFCC)=9$.
            \item[{\textbf{(ii).}}]{\textbf{The gauge fixing conditions}}\\
                Once the number of independent first class constraints has been determined,
                the issue of finding the number of independent gauge fixing conditions is
                straightforward. It can be shown \cite{Teit},\cite{Henn} that for a gauge
                theory that obeys the Dirac conjecture, the number of such independent
                gauge fixing conditions is in fact necessarily identical to the number of
                independent first class constraints. Hence, for the particular case of the
                $\Sigma\Phi EA$ theory $(\#IGC)=(\#IFCC)=9$.
        \end{enumerate}
    Introducing the number of independent first class constraints and gauge
    conditions determined above into (\ref{DOFeqn}), the revised calculation
    yields $(\#PDOF)=0$, which means that the $\Sigma\Phi EA$ theory has no
    local physical degrees of freedom. Of course, this doesn't mean that the
    theory is necessarily trivial. It only means that it is a topological field
    theory, and like any other such topological theory it is locally trivial,
    while globally it can still have non-trivial physical degrees of freedom
    depending on the topology of the spacetime manifold.

\section{The classical sectors of the $\Sigma\Phi EA$ theory}

\indent
    Having established that the $\Sigma\Phi EA$ theory is a topological field theory,
    the next logical step is to determine and classify the distinct gauge-equivalent
    classes of solutions of the theory. For topological theories, since any such theory
    is locally trivial, any such classification is usually based on the analysis
    of global observables, or in other words, on the existence of global gauge invariant
    quantities, and in the general case the classification will be related to the
    existence of the Casimir invariants of the gauge algebra, and/or of other quantities
    that are constant along certain gauge orbits \cite{Freidel1}. Therefore, in order to
    make such a classification possible, a natural approach would be to first determine a
    complete set of such global observables.\\
\indent
    However, compared to the case of the $BCEA$ model, constructing global observables for
    the $\Sigma\Phi EA$ theory is even in principle a highly nontrivial task. This is
    mainly due to the fact that the latter theory contains scalar and tensorial
    matter fields, and as such - to the best knowledge of the authors - it cannot
    be written as either a BF theory or as a Chern-Simons theory\footnote{As will be
    shown in Section VII, the $\Sigma\Phi EA$ theory can be written as a generalized
    Chern-Simons theory involving a multiform connection. Unfortunately, such a
    multiform formulation of the theory makes it even more difficult to construct
    global observables.}. Under these circumstances, none of the more traditional
    techniques are available in the $\Sigma\Phi EA$ theory, and one must look
    elsewhere for the construction of such global observables\footnote{The
    issue of global observables in the $\Sigma\Phi EA$ theory is currently under
    under study, and the results will be presented in a companion paper}.\\
\indent
    Nevertheless, it is still possible to develop a classification of the classical
    sectors of the $\Sigma\Phi EA$ theory if one makes the essential observation
    that one can construct a very simple gauge invariant quantity using
    the fields $\Phi^{i}$ exclusively. This quantity is $\Phi^{i}\Phi_{i}$, and it is straightforward
    to check its gauge invariance by directly using the gauge transformations
    (\ref{eqngauge1}). Furthermore, one can also observe that the fields $\Phi$
    transform non-trivially only under Lorentz transformations, and therefore we can
    view these fields as a Minkowskian "vector" $\Phi=(\Phi^{i})$, whose magnitude
    $\Phi\cdot\Phi=\Phi^{i}\Phi_{i}$ is left invariant under the action of $SO(2,1)$.\\
\indent
    It follows from the above considerations that the simplest criterion for classifying
    the solutions of the $\Sigma\Phi EA$ theory should be based on the values of the
    invariant quantity $\Phi\cdot\Phi=\Phi^{i}\Phi_{i}$. As a Minkowskian vector, $\Phi$
    can be timelike, spacelike, null or identically zero, corresponding to a magnitude
    $\Phi^{i}\Phi_{i}$ that is negative, positive or zero. The zero value of the magnitude
    is degenerate, in the sense that it contains the cases where $\Phi$ is null or
    identically zero, and in the null case, additional degeneracy arises from the existence
    of an extra parameter that specifies whether $\Phi$ is a futurelike or pastlike null
    vector. We will discuss each of these cases separately.
        \begin{enumerate}
            \item[a.] The case $\Phi\equiv 0$\\
                In this case, the equations of motion (\ref{eqnmotn1}) reduce to:
                    \be
                        R^{i}[A]=0 ;\; DE^{i}=0 ;\; D\Sigma^{i}=0 \label{eqnmotna}
                    \ee
                and the classification of solutions can be further split as follows:
                    \begin{itemize}
                        \item[a1)] If the fields $\Sigma$ and $A$ are parallel, i.e. if
                            $\epsilon^{ijk}A_{j}\wedge\Sigma_{k}=0$, then the dynamics of
                            the 2-form fields $\Sigma$ decouples from the dynamics of
                            the triad and the spin connection, and we recover pure gravity
                            in (2+1) dimensions in an arbitrary background field $\Sigma$.
                        \item[a2)] If the fields $\Sigma$ and $A$ are not parallel, i.e. if
                            $\epsilon^{ijk}A_{j}\wedge\Sigma_{k}\neq0$, the dynamics of
                            the 2-form field cannot be decoupled from the dynamics of gravity
                            anymore, and the solution will contain three sets of dynamically
                            intracting fields.
                    \end{itemize}
                                               \item[b.] The case $\Phi$ is timelike.\\
                In this case the field $\Phi$ can be put in the form $\Phi=(\Phi^{0},0,0)$,
                and it follows immediately from the equations of motion (\ref{eqnmotn1})
                that in fact $\Phi^{0}$ must be a constant and $A_{1}=A_{2}=0$.
            \item[c.] The case $\Phi$ is spacelike.\\
                This case is very similar to the timelike case. The field $\Phi$ can be
                put in the form $\Phi=(0,\Phi^{1},0)$, and it follows immediately from
                the equations of motion (\ref{eqnmotn1}) that in fact $\Phi^{1}$ must be
                a constant and $A_{0}=A_{2}=0$
            \item[d.] The case $\Phi$ is null.\\
                As mentioned earlier, in this case, the classification of the orbits can
                be further split based on whether the field $\Phi$ is futurelike or
                pastlike. Since the analysis of the solutions in the two cases is very
                similar, we will only consider the case where $\Phi$ is pastlike null with
                the form $\Phi=(-\phi,\phi,0)$. Under these circumstances, it follows from
                the field equations (\ref{eqnmotn1}) that $\phi$ is a constant field,
                and the spin connection fields are such that $A^{2}=0$ and $A^{0}=-A^{1}$.
                Introducing the notation $\tilde{A}=A^{0}$, $\tilde{E}=E^{0}+E^{1}$,
                $\tilde{\Sigma}=\Sigma^{0}+\Sigma{1}$, the equations of motion
                (\ref{eqnmotn1}) now become:
                    \bea
                        d\tilde{A}=0; \; d\tilde{E}=0; \;d\tilde{\Sigma}=0 \nonumber\\
                        \tilde{A}\wedge E^{2}-\Sigma^{2}\phi=0 \nonumber\\
                        \tilde{A}\wedge \Sigma{2}=0 \label{eqnsmotnd}\\
                        dE^{2}+\tilde{A}\wedge \tilde{E}+
                                        \tilde{\Sigma}\phi=0\nonumber\\
                        d\Sigma^{2}-\tilde{A}\wedge\tilde{\Sigma}=0\nonumber
                    \eea
        \end{enumerate}

\section{Examples of solutions}

\subsection{Black-hole solution of the $\Sigma\Phi EA$ Model}

\indent
    In this section, we show that the $\Sigma\Phi EA$ model admits
    the BTZ black-hole geometry as a solution, and we calculate
    the the conserved Noether charges associated with this
    solution. For more details regarding the BTZ black-hole,
    the reader should consult \cite{btz1}, \cite{btz2}.

\subsubsection{The BTZ black-hole solution}

\indent
    The BTZ black-hole geometry can be described by the triad fields \cite{bceabtz},
    \cite{btz2}:
        \bea
            E^{0}&=& \sqrt{\nu^{2}(r)-1}(\frac{r_{+}}{\textit{l}}dt-r_{-}d\phi)\nonumber\\
            E^{1}&=& \frac{\textit{l}}{\nu(r)}d[\sqrt{\nu^{2}(r)-1}]\label{btztriad}\\
            E^{2}&=& \nu(r)(-\frac{r_{-}}{\textit{l}}dt+r_{+}d\phi)\nonumber
        \eea
    where
        \bea
            r^{2}_{+}&=& \frac{\textit{M}\textit{l}^2}{2}\,\{ 1+
                        \sqrt{1-(\textit{J}/\textit{M}\textit{l})^{2}}\} \nonumber\\
            r^2_-&=&\frac{\textit{M}\textit{l}^2}{2}\,\{ 1-
                        \sqrt{1-(\textit{J}/\textit{M}\textit{l})^{2}}\} \label{horradii}
        \eea
    are the outer and respectively inner horizon radii, satisfying
    $r_+r_-= \textit{J}\textit{l}/2$, the function $\nu(r)$ is given by the expression:
        \be
            \nu^{2}(r)=\frac{r^{2}-r^{2}_{-}}{r^{2}_{+}-r^{2}_{-}}
        \ee
    and the parameters $\textit{M}$, $\textit{J}$ and $\textit{l}$
    have the same significance as described in Section II.\\
\indent
    In order to find the matter fields of the $\Sigma\Phi EA$
    theory associated with the the geometry of the black hole,
    one has to solve the equations of motion (\ref{eqnmotn1}) with
    the triad fields given by (\ref{btztriad}). For simplicity
    we will solve the equations of motion in the gauge $A^{i}=0$.
    With appropriate rescaling, a class of matter fields that can
    be obtained in this way is given by:
        \bea
            \vec{\Phi}&=&(0,1,0)\nonumber\\
            \Sigma^{0}&=&\frac{r}{2\beta \sqrt{r^{2}-r^{2}_{+}}}\,
                       \{\frac{r_{+}}{\textit{l}}dr\wedge dt-
                       r_{-}dr\wedge d\phi\}\label{sfea_flds}\\
            \Sigma^{1}&=&\textit{arbitrary closed 1-form}\nonumber\\
            \Sigma^{2}&=&\frac{r}{2\beta \sqrt{r^{2}-r^{2}_{-}}}\,
                       \{-\frac{r_{-}}{\textit{l}}dr\wedge dt-
                       r_{+}dr\wedge d\phi\}\nonumber
        \eea
\indent
    As can be immediately observed from (\ref{sfea_flds}) the
    black-hole solution in the $\Sigma\Phi EA$ model differs
    significantly from the corresponding solution in the BCEA
    theory, since in the latter case there is no arbitrariness
    in the matter fields once the gauge for
    the spin connection coefficients has been fixed. In other
    words, the black-hole solution has additional gauge freedom in
    the $\Sigma\Phi EA$ theory as compared to its BCEA
    counterpart, and this additional gauge freedom is directly
    related to the 0-form/2-form choice for the matter fields.

\subsubsection{The Noether charges for the BTZ black-hole solution}

\indent
    Since the $\Sigma\Phi EA$ theory is a topological theory,
    its diffeomorphism invariance is equivalent on-shell
    with invariance under the infinitesimal gauge transformations
    (\ref{eqngauge1}). Hence one can use the formalism developed
    in \cite{bceabtz} to calculate the Noether charges associated
    with its symmetries.\\
\indent
    We begin by first summarizing the formalism in \cite{bceabtz}.
    Assuming the Lagrangian density to be a functional $\textbf{L}[\beta]$
    of generic fields $\beta$, under a first order arbitrary variation
    of these fields the corresponding variation of the Lagrangian density
    can be written as:
        \be
            \delta\textbf{L}[\beta]=d\Theta[\beta,\delta\beta]\label{fovar}
        \ee
    where in writing (\ref{fovar}) we have already taken the equations
    of motion into account. The 2-form (confining ourselves to the
    3-dimensional case) $\Theta[\beta,\delta\beta]$
    appearing in the RHS of (\ref{fovar}) is called the symplectic
    potential current density and will play a major role in the
    construction of the symplectic structure of the theory.
    Similarly, under a  symmetry transformation of the fields
    $\delta_{g}\beta$, where g is an element of the symmetry
    group G, the invariance of the Lagrangian density can be expressed as:
        \be
            \delta_{g}\textbf{L}[\beta]=d\alpha[\beta,\delta_{g}\beta]\label{symvar}
        \ee
    where now $\alpha$ is some arbitrary 2-form.
    Using now the two forms $\Theta$ and $\alpha$, we can
    construct the 2-form:
        \be
            \textbf{j}[g]=\Theta[\beta,\delta_{g}\beta]-
                            \alpha[\beta,\delta_{g}\beta]\label{dualcur}
        \ee
    and it is clear that this 2-form is closed when the equations
    of motion are satisfied. The 2-form $\textbf{j}[g]$
    is nothing other than the Hodge dual of the Noether current
    associated with the symmetry generated by the symmetry
    group element $g\in G$, and its integral over a Cauchy surface
    $\cal{C}$ yields the conserved charges $q[g]$ associated
    associated with the symmetry generated by g. Furthermore, if
    $\textbf{j}[g]$ is also exact then $\textbf{j}[g]=d\textbf{Q}[g]$,
    and the surface integral over $\cal{C}$ reduces to a line
    integral over $\partial\cal{C}$.\\
\indent
    Referring now to the case of diffeomorphism invariance,
    the symmetry transformation of the fields in this case is given
    by $\delta_{\xi}\beta=\cal{L}_{\xi}\beta$, where $\cal{L}$ is
    the Lie derivative operator, and $\xi$ is the diffeomorphism
    generating vector field. Under these circumstances it can be
    shown that the dual Noether current (\ref{dualcur}) can be put
    in the form:
        \be
            \textbf{j}[\xi]=\Theta[\beta,\cal{L}_{\xi}\beta]-
                                \xi\cdot\textbf{L}\label{difcur}
        \ee
    where the dot in (\ref{difcur}) denotes the contraction of the
    vector filed $\xi$ with the first index of the Lagrangian
    density 3-form. For the particular case of diffeomorphisms
    generated by asymptotic time translations $t^{\mu}$ and by
    asymptotic rotations $\varphi^{\mu}$, it has been shown
    \cite{Wald1}, \cite{Wald2}, \cite{Wald3}, that the
    corresponding conserved charges, i.e. the canonical energy and
    the canonical angular momentum are given by the line integrals
    along a circle at constant time and infinite radius according to
    the relations:
        \bea
            \cal{E}&=& \int_{\infty}(\textbf{Q}[t]-t\cdot G)\nonumber\\
            \cal{J}&=& \int_{\infty}\textbf{Q}[\varphi]\label{nochrgs}
        \eea
    provided one can determine the 2-form G from the condition:
        \be
            \delta_{0}\int_{\infty}t\cdot G = \int_{\infty}t\cdot
                                    \Theta[\beta,\delta_{0}\beta]\label{Gcond}
        \ee
    where $\delta_{0}$ are variations of the fields within the space of
    solutions of the theory.\\
\indent
    Specializing now to the case of the BTZ black-hole solution in
    the $\Sigma\Phi EA$ model, it is straightforward to show that
    the symplectic potential current density $\Theta$ is given by
    the expression:
        \be
            \Theta[\Sigma,E,\delta A,\delta \Phi]=\Sigma^{i}\wedge\delta\Phi_{i}
                            -E^{i}\wedge\delta A_{i}\label{sfeatheta}
        \ee
    and since on-shell the Lagrangian of the theory is obviously
    zero, the Noether dual current for the case of diffeomorphisms
    is simply:
        \be
            \textbf{j}[\xi]=\Sigma^{i}\wedge{\cal{L}}_{\xi}\Phi_{i}
                            -E^{i}\wedge{\cal{L}}_{\xi}A_{i}\label{sfeacur}
        \ee
    and the conserved Noether charges in (\ref{nochrgs}) are determined
    exclusively by the 1-form $\textbf{Q}$.\\
\indent
    On-shell, the diffeomeorphisms are equivalent to the gauge
    transformations (\ref{eqngauge1}), which enables us to write:
        \bea
            {\cal{L}}_{\xi}\Phi_{i}&=&\epsilon_{ijk}\Phi^{j}\tau^{k}\nonumber\\
            {\cal{L}}_{\xi}A_{i}&=&D\tau_{i}\label{onshelldiff}
        \eea
    where $\tau^{i}=\xi\cdot A^{i}$, and with (\ref{onshelldiff}),
    it is straightforward to show that if the equations of motion
    are satisfied, the dual Noether current (\ref{sfeacur})
    becomes simply:
        \be
            \textbf{j}[\tau(\xi)]=d(E^{i}\tau_{i})\label{sfeacur1}
        \ee
    Under these circumstances the 1-form $\textbf{Q}$ is given by
    the expression:
        \be
            \textbf{Q}=E^{i}\tau_{i}\label{qcharge}
        \ee
    and since the gauge parameters $\tau^{i}$ are proportional to the to
    the connection fields $A^{i}$, it is clear that for the BTZ
    black-hole solution $\tau^{i}=0$, which in turn means that $\textbf{Q}=0$,
    and hence the conserved Noether charges vanish identically:
        \be
            {\cal{M}}_{\Sigma\Phi EA}={\cal{J}}_{\Sigma\Phi EA}=0 \label{sfeabtzchrgs}
        \ee
\indent
    This result is extremely interesting, especially if we correlate it with
    the corresponding result for the BTZ black-hole solution in the BCEA model. While we
    do not have a more fundamental understanding of this effect, a possible explanation
    could be that the Noether charges and correspondingly the original gravitational
    fields of the black-hole are "screened" (to be thought of in a "generalized"
    sense) by the topological matter fields, very much like electric charges are being
    screened by other configurations of charges or alternatively, in the macroscopic
    sense, like electromagnetic fields being screened (even up to extinction) by matter.
    However, such an explanation suffers from the obvious drawback that such screening
    of charges requires charges of opposite signs, and while an opposite angular momentum
    is a realistic interpretation, an "opposite" mass/energy is a rather unacceptable
    concept.

\subsection{Cosmological solutions of the $\Sigma\Phi EA$ model}

\indent
    A very particular feature of the $\Sigma\Phi EA$ model is that
    for the sector of the theory for which $\Phi$ is
    timelike, the model admits cosmological solutions of the
    Friedmann-Robertson-Walker type. To the best of our knowledge,
    this is the first  class of topological matter models in (2+1)
    dimensions that exhibits such characteristics.

    Consider the equations of motion (\ref{eqnmotn1}) for the model. In the
    gauge $A^{i}=0$, if $\Phi$ is constant and pure timelike, they reduce to:
        \bea
            dE^{0}&=&0\nonumber\\
            dE^{1}-\Sigma^{2}\Phi^{0}&=&0\nonumber\\
            dE^{2}+\Sigma^{1}\Phi^{0}&=&0\label{eqn39}\\
            d\Sigma^{i}&=&0\nonumber
        \eea
    and consider a (2+1)-dimensional Friedman-Robertson-Walker type of
    metric:
        \be
            ds^{2}=-dt^{2}+f(t)dx^2\label{eqn40}
        \ee
    where $dx^{2}$ is a 2-dimensional spatial metric, and $f(t)$ is an arbitrary
    function of time. For such a metric, we can always choose $E^{0}=1$, in which
    case the first of the equations of motion in (\ref{eqn39}) is satisfied identically.
    Furthermore, for any well behaved 2-dimensional spatial metric that allows us to
    determine $E^{1}$ and $E^{2}$ in a closed and convenient form, it is obvious from
    the rest of the equations of motion in (\ref{eqn39}) that we can also find two
    fields $\Sigma^{1}$ and $\Sigma^{2}$ such that these two equations of motion are also
    satisfied. These latter two fields will be obviously given by the expressions:
        \bea
            \Sigma^{1}&=&-\frac{dE^{2}}{\Phi^{0}}\nonumber\\
            \Sigma^{2}&=&\frac{dE^{1}}{\Phi^{0}}\label{eqn41}
        \eea

\section{The global charges of the $\Sigma\Phi EA$ model}

\indent
    Besides the gauge invariant observables in the bulk that can be
    constructed from the fields of the theory, there is another
 class of observables  that are associated with the boundaries of
    the spacelike surface S. In the dedicated terminology, these
    observables are called global charges, and they arise from the
    requirement that the symmetry generators of the theory be
    differentiable. In the following, we will analyze these global
    charges, and the algebra they generate.\\
\indent
    Before proceeding with the analysis, it is necessary to make
    some remarks regarding the boundaries of the spacelike surface
    $S$. In the general case the boundary of $S$ can consist of several
    disconnected components, which can be internal (e.g. the horizon
    of a black hole) or external (e.g. asymptotic boundaries at spatial
    infinity). For the time being however, we will restrict ourselves
    only to the case where the spacelike surface $S$ has a single boundary,
    which will be considered to have the topology of a circle, and we will
    make no distinction of whether the boundary is internal or external.
    The generalization to multiple disconnected components is
    straightforward, and we will specialize the analysis to each
    type of boundary - internal or external - in the appropriate
    context.

\subsection{Field independent gauge parameters}

\indent
    Consider once again the constraints (\ref{constraints}) of the
    $\Sigma\Phi EA$ model. For reasons that will become clear below, it
    is more convenient at this time to revert to the notation
    $K^{iA}$ for the constraints $Q^{i}, R^{i}$ and to write the
    constraints in the form:
        \bea
            J^{i}&=&D_{[1}E^{i}_{2]}-
                    2\,\epsilon^{ijk}\Sigma_{j12}\Phi_{k}\approx 0 \nonumber\\
            P^{i}&=&\partial_{[1}A^{i}_{2]}+
                    \frac{1}{2}\epsilon^{ijk}A_{j[1}A_{k2]}\approx 0 \label{eqn1}\\
            K^{i}_{A}&=&D_{A}\Phi^{i}\approx 0 \nonumber
        \eea
    where in (\ref{eqn1}) we have used the notation $D_A=\partial_{A}+\epsilon^{ijk}A_{jA}$\\
\indent
    With these constraints we can construct the following three types of
    smeared gauge symmetry generators:
        \bea
            G_{J}[\alpha]&=&\int_{S}d^{2}x\alpha_{i}J^{i}\nonumber\\
            G_{P}[\beta]&=&\int_{S}d^{2}x\beta_{i}P^{i}\label{eqn2}\\
            G_{K}[\gamma]&=&\int_{S}d^{2}x\gamma_{i}\wedge K^{i}\nonumber
        \eea
    where $\alpha_{i}$, $\beta_{i}$ and $\gamma_{i}$ are 0-form and respectively
    1-form field-independent gauge parameters on the spacelike surface $S$. As
    defined above however, these generators are not differentiable, and it is
    straightforward to show that under a variation of the fields, the variation
    of these generators contains boundary terms that preclude their differentiability:
        \bea
            \delta G_{J}[\alpha]&=&\int_{S}d^{2}x\Big\{-\epsilon^{AB}\big[(D_{A}\alpha_{i})(\delta E^{i}_{B})+
                    \epsilon^{ijk}[\alpha_{i}(\delta A_{jA})E_{kB}-{}\nonumber\\
                    &&{}-2\alpha_{i}(\delta\Sigma_{j12})\Phi_{k}-
                    2\alpha_{i}\Sigma_{j12}(\delta\Phi_{k})]\big]\Big\}+
                    \int_{\partial S}dx^{A}\alpha_{i}(\delta E^{i}_{A})\nonumber\\
            \delta G_{P}[\beta]&=&\int_{S}d^{2}x[\epsilon^{AB}(D_{A}\beta_{i})(\delta A^{i}_{B})]+
                    \int_{\partial S}dx^{A}\beta_{i}(\delta A^{i}_{A})\label{eqn3}\\
            \delta G_{K}[\gamma]&=&\int_{S}d^{2}x\big\{\epsilon^{AB}[(D_{A}\gamma_{iB})(\delta\Phi^{i})-
                    \epsilon^{ijk}\gamma_{iA}(\delta A_{jB})\Phi_{k}]\big\}-{}\nonumber\\
                    &&{}-\int_{\partial S}dx^{A}\gamma_{iA}(\delta\Phi^{i})\nonumber
        \eea
    where $\partial S$ is the boundary of $S$. \\
\indent
    The most straightforward way to to make these generators differentiable would be,
    of course, to simply add to each of the variations in (\ref{eqn3}) a boundary term
    [infinitesimal(global) charge] that cancels the already existing one, i.e. to add
    to each of the $\delta G_{J}[\alpha]$, $\delta G_{P}[\beta]$, $\delta G_{K}[\gamma]$
    the following corresponding terms:
       \bea
            \delta Q_{J}[\alpha]&=& -\int_{\partial S}dx^{A}\alpha_{i}(\delta E^{i}_{A})\nonumber\\
            \delta Q_{P}[\beta]&=& -\int_{\partial S}dx^{A}\beta_{i}(\delta A^{i}_{A})\label{eqn4}\\
            \delta Q_{K}[\gamma]&=& \int_{\partial S}dx^{A}\gamma_{iA}(\delta\Phi^{i})\nonumber
        \eea
    In the general case, this approach does not entirely solve the differentiability
    problem of the gauge symmetry generators unless the infinitesimal charges in
    (\ref{eqn4}) are integrable. For the case of field-independent gauge parameters
    however, the infinitesimal charges in (\ref{eqn4}) can be integrated straightforwardly
    to yield (up to integration constants which can be chosen to vanish):
        \bea
            Q_{J}[\alpha]&=&-\int_{\partial S}dx^{A}\alpha_{i}E^{i}_{A}\nonumber\\
            Q_{P}[\beta]&=&-\int_{\partial S}dx^{A}\beta_{i}A^{i}_{A}\label{eqn7}\\
            Q_{K}[\gamma]&=&\int_{\partial S}dx^{A}\gamma_{iA}\Phi^{i}\nonumber
        \eea
    and consequently, with the global charges in (\ref{eqn7}), one can define a set
    of differentiable smeared gauge symmetry generators through the relations:
        \bea
            \tilde{G}_{J}[\alpha]&=&G_{J}[\alpha]+Q_{J}[\alpha]\nonumber\\
            \tilde{G}_{P}[\beta]&=&G_{P}[\beta]+Q_{P}[\beta]\label{eqn5}\\
            \tilde{G}_{K}[\gamma]&=&G_{K}[\gamma]+Q_{K}[\gamma]\nonumber
        \eea
    These differentiable generators have now well-defined Poisson
    brackets with themselves and with any other differentiable
    functional of the fields, and it is a simple exercise to show
    that they generate the infinitesimal gauge transformations
    (\ref{eqngauge1}).\\
\indent
    The next necessary step in determining the algebra of global charges
    is to calculate the Poisson brackets of the generators in (\ref{eqn5})
    with themselves. Since the canonically conjugate variables of the theory
    are $(A,E)$ and $(\Sigma,\Phi)$, a straightforward calculation yields only
    three non-trivial such brackets:
        \bea
            \big\{\tilde{G}_{J}[\alpha],\tilde{G}_{J}[\tau]\big\}_{PB}&=&
                \int_{S}d^{2}x[\epsilon_{imn}\alpha^{i}\tau^{m}J^{n}]-
                \int_{\partial S}dx^{A}\epsilon_{imn}\alpha_{i}\tau^{m}E^{n}_{A}\nonumber\\
            \big\{\tilde{G}_{J}[\alpha],\tilde{G}_{P}[\beta]\big\}_{PB}&=&
                \int_{S}d^{2}x[\epsilon_{imn}\alpha^{i}\beta^{m}P^{n}]+
                \int_{\partial S}dx^{a}\alpha_{i}(D_{a}\beta^{i}) \label{eqn6}\\
            \big\{\tilde{G}_{J}[\alpha],\tilde{G}_{K}[\gamma]\big\}_{PB}&=&
                \int_{S}d^{2}x[\epsilon_{imn}\alpha^{i}\gamma^{m}_{1}K^{n}_{2}-
                \epsilon_{imn}\alpha^{i}\gamma^{m}_{2}K^{n}_{1}]+
                \int_{\partial S}dx^{A}\epsilon_{imn}\alpha_{i}\gamma^{m}_{A}\Phi^{n}\nonumber
        \eea
    Comparing now the boundary terms in (\ref{eqn6}) with the
    expression of the global charges in (\ref{eqn7}), one can
    immediately see that the algebra of the differential
    generators closes under the Poisson bracket, and can be
    rewritten more compactly in the form:
        \bea
            \big\{\tilde{G}_{J}[\alpha],\tilde{G}_{J}[\tau]\big\}_{PB}&=&
                \tilde{G}_{J}\big[[\alpha, \tau]\big]\nonumber\\
            \big\{\tilde{G}_{J}[\alpha],\tilde{G}_{P}[\beta]\big\}_{PB}&=&
                \tilde{G}_{P}\big[[\alpha, \beta]\big]+
                \int_{\partial S}dx^{A}\alpha_{i}(\partial_{A}\beta^{i})\label{eqn8}\\
            \big\{\tilde{G}_{J}[\alpha],\tilde{G}_{K}[\gamma]\big\}_{PB}&=&
                \tilde{G}_{K}\big[[\beta, \gamma]\big]\nonumber
        \eea
    where in (\ref{eqn8}) we have used the notation
    $[\alpha,\beta]\equiv \epsilon_{ijk}\alpha^{i}\beta^{j}$. Since the boundary
    term still remaining in (\ref{eqn8}) is independent of the fields, the above
    algebra can be interpreted as some sort of central extension of an infinite
    dimensional version of a Poincar\'{e} algebra inhomogeneized by an additional set
    of translations. Alternatively, it can also be viewed as the
    central extension of a Ka\v{c}-Moody algebra inhomogeneized by
    two (infinite) sets of abelian ``translation" generators.\\
\indent
    Having cast the algebra of the differentiable generators into a closed form,
    the global charges will obey the same algebra, with the only difference that
    the Poisson brackets are replaced by the corresponding Dirac brackets
    \cite{Banados1}. Therefore, the Dirac algebra of the
    charges will be given by the relations:
        \bea
            \big\{Q_{J}[\alpha],Q_{J}[\tau]\big\}_{D}&=&
                Q_{J}\big[[\alpha, \tau]\big]\nonumber\\
            \big\{Q_{J}[\alpha],Q_{P}[\beta]\big\}_{D}&=&
                Q_{P}\big[[\alpha, \beta]\big]+
                \int_{\partial S}dx^{a}\alpha_{i}(\partial_{a}\beta^{i})\label{eqn9}\\
            \big\{Q_{J}[\alpha],Q_{K}[\gamma]\big\}_{D}&=&
                Q_{K}\big[[\alpha, \gamma]\big]\nonumber
        \eea
    with all the rest of the Dirac brackets vanishing.\\
\indent
    So far, we have described the algebra of the smeared constraints and the
    corresponding algebra of of global gauge charges for the theory formulated
    on the Lorentz group/algebra, i.e. for the theory whose action is defined
    as the trace over antisymmetric products of Lorentz algebra-valued forms.
    It is possible, however (and also convenient for the discussion of the global
    diffeomorphism charges, as it will become clear in the next section), to
    reformulate the theory as a theory on the Poincar\'{e} group/algebra, and discuss
    the global charges within this new framework.\\
\indent
    In order to reformulate the theory with the Poincar\'{e} group/algebra, one
    must first note that the pure gravity term in the $\Sigma\Phi EA$ action
    (\ref{ssfea}) can be written, up to surface terms, as a Chern-Simons action
    with the Poincar\'{e} connection:
        \be
            {\cal{A}}=A_{i}\bar{J}^{i}+E_{i}\bar{P}^{i}={\cal{A}}_{a}T^{a}\label{eqn10}
        \ee
    where latin indices from the beginning of the alphabet $\{a,...,h\}$ are
    Poincar\'{e} Lie algebra indices taking the values $\{0,...,5\}$, and
    $\{T^{a}\}=\{\bar{J}^{0},...,\bar{P}^{2}\}$ are the generators of the
    Poincar\'{e} algebra. Defining now the Poincar\'{e} matter fields:
        \bea
            \Sigma=\Sigma_{i}\bar{J}^{i}=\Sigma_{a}T^{a} \nonumber\\
            \Phi=\Phi_{i}\bar{P}^{i}=\Phi_{a}T^{a}\label{eqn11}
        \eea
    and the Poincar\'{e} covariant derivative as:
        \be
            \tilde{D}\Phi=d\Phi+[{\cal{A}},\Phi]\label{eqn12}
        \ee
    the action (\ref{ssfea}) can be rewritten as a Chern-Simons action with
    the Poincar\'{e} connection (\ref{eqn10}) plus a Poincar\'{e} topological
    matter term:
        \be
            S[\Sigma,\Phi,E,A]=\int_{M}\tilde{Tr}\Big[\frac{1}{2}{\cal{A}}d{\cal{A}}+
                \frac{1}{3}{\cal{A}}^{3}+\Sigma\wedge\tilde{D}\Phi\Big]\label{eqn13}
        \ee
    where in (\ref{eqn13}) the wedge product of forms in the
    Chern-Simons terms is implicitly assumed, and the trace$\tilde{Tr}$ is the
    non-degenerate invariant bilinear form on the Poincar\'{e}
    algebra defined in terms of the Lorentz algebra generators as:
        \be
            \tilde{Tr}(\bar{J}^{i}\bar{P}^{j})=\eta^{ij},\;
            \tilde{Tr}(\bar{J}^{i}\bar{J}^{j})=0,\;
            \tilde{Tr}(\bar{P}^{i}\bar{P}^{j})=0.\label{eqn14}
        \ee
\indent
    Within the Poincar\'{e} formulation, and using the notations developed for the
    Lorentz formulation of the theory, one can define a Poincar\'{e} constraint:
        \be
            G=P_{i}\bar{J}^{i}+J_{i}\bar{P}^{i}=G_{a}T^{a}\label{eqn15}
        \ee
    together with a Poincar\'{e} gauge parameter:
        \be
            \lambda=\alpha_{i}\bar{J}^{i}+\beta_{i}\bar{P}^{i}=\lambda_{a}T^{a}\label{eqn16}
        \ee
    Consequently, using these two quantities, one can define a Poincar\'{e}
    symmetry generator:
        \be
            G_{JP}[\lambda]=\int_{S}d^{2}x\lambda_{a}G^{a}\label{eqn17}
        \ee
    and it is a matter of straightforward calculation to show
    that:
        \be
            G_{JP}[\lambda]=G_{J}[\alpha]+G_{P}[\beta]\label{eqn18}
        \ee
\indent
    In a similar manner, by defining the Poincar\'{e} 1-form constraint:
        \be
            K=K_{i}\bar{P}^{i}=K_{a}T^{a}\label{eqn19}
        \ee
    and the Poincar\'{e} 1-form gauge parameter:
        \be
            \gamma=\gamma_{i}\bar{J}^{i}=\gamma_{a}T^{a}\label{eqn20}
        \ee
    the gauge symmetry generator $G_{K}[\gamma]$ can be rewritten
    as a Poincar\'{e} symmetry generator:
        \be
            G_{K}[\gamma]=\int_{s}d^{2}x\gamma_{i}\wedge K^{i}
                         =\int_{s}d^{2}x\gamma_{a}\wedge K^{a}\label{eqn21}
        \ee
    where, for obvious reasons we are using the same notation for
    this generator in both formulations.\\
\indent
    Having the two Poincar\'{e} symmetry generators above, we can repeat or
    translate the previous analysis regarding their differentiability. For
    the case of field-independent gauge parameters $\lambda$,
    $\gamma$, the differentiability of these generators is ensured
    by adding to them corresponding global charges:
        \bea
            Q_{JP}[\lambda]&=&-\int_{\partial S}dx^{A}\lambda_{a}{\cal{A}}^{a}_{A}\nonumber\\
            Q_{K}[\gamma]&=&\int_{\partial S}dx^{A}\gamma^{a}_{A}\Phi_{a}\label{eqn22}
        \eea
    and consequently one can define differentiable gauge symmetry
    generators:
        \bea
            \tilde{G}_{JP}[\lambda]&=&G_{JP}[\lambda]+Q_{JP}[\lambda]\nonumber\\
            \tilde{G}_{K}[\gamma]&=&G_{K}[\gamma]+Q_{K}[\gamma]\label{eqn23}
        \eea
    which now have well-defined Poisson brackets with themselves and with any
    other differentiable functional of the fields.\\
\indent
    The Poisson algebra of these generators can be straightforwardly calculated
    as:
        \bea
            \big\{\tilde{G}_{JP}[\lambda],\tilde{G}_{JP}[\eta]\big\}_{PB}&=&
                \tilde{G}_{JP}\big[[\lambda,\eta]\big]-
                \int_{\partial S}dx^{A}\lambda_{a}(\partial_{A}\eta^{a})\nonumber\\
            \big\{\tilde{G}_{JP}[\lambda],\tilde{G}_{K}[\gamma]\big\}_{PB}&=&
                \tilde{G}_{K}\big[[\lambda, \beta]\big]\label{eqn24}\\
            \big\{\tilde{G}_{K}[\gamma],\tilde{G}_{K}[\bar{\gamma}]\big\}_{PB}&=&0\nonumber
        \eea
    which in turn yields the Dirac algebra of global charges:
        \bea
            \big\{Q_{JP}[\lambda],Q_{JP}[\eta]\big\}_{D}&=&Q_{JP}\big[[\lambda,\eta]\big]-
                \int_{\partial S}dx^{A}\lambda_{a}(\partial_{A}\eta^{a})\nonumber\\
            \big\{Q_{JP}[\lambda],Q_{K}[\gamma]\big\}_{D}&=&
                Q_{K}\big[[\lambda,\gamma]\big]\label{eqn25}\\
            \big\{Q_{K}[\gamma],Q_{K}[\bar{\gamma}]\big\}_{D}&=&0\nonumber
        \eea
    where in (\ref{eqn24},\ref{eqn25}), the commutator of gauge parameters stands for
    $[\lambda,\eta]=f_{abc}\lambda^{a}\eta^{b}$, with $f_{abc}$ the structure
    constants of the Poincar\'{e} algebra.\\
\indent
    It is clear now, in the Poincar\'{e} formulation, that the  Dirac algebra
    of global charges is an inhomogenization of the Ka\v{c}-Moody algebra of
    charges (with a central term) for pure gravity. Indeed, if we "turn off"
    the matter fields, and consequently the symmetry generator $\tilde{G}_{K}$,
    we are only left with the first Poisson bracket in (\ref{eqn24}) and
    respectively with the first Dirac bracket in (\ref{eqn25}), and the latter
    can be recognized once again as the algebra of global gauge charges for gravity
    in (2+1) dimensions \cite{Banados1}. Furthermore, it is also clear from the
    form of the Dirac algebra of charges, and in fact also from the Poisson algebra
    of the gauge symmetry generators, that the inhomogeneization of the respective
    algebras is of Poincar\'{e} type (semi-direct product type), i.e. the Lorentz-like
    algebra with central charge is inhomogenized by a set of Poincar\'{e}-like abelian
    translations.\\
\indent
    In order to better illustrate the above considerations, and also in order
    to put the algebra of charges (\ref{eqn25}) in a form that is more amenable
    to the traditional Dirac quantization procedure it is useful to consider
    the Fourier modes of the free fields on the boundary $\partial S$. Since
    these fields are considered to be periodic on the boundary (which in the
    following will be assumed to be a circle with the periodic coordinate
    $\varphi$) they admit the following Fourier series expansion:
        \bea
            {\cal{A}}^{a}_{\varphi}&=&
                    \sum_{n=-\infty}^{n=\infty}B^{a}_{n}\,e^{in\varphi}\nonumber\\
            \Phi^{a}&=&
                    \sum_{n=-\infty}^{n=\infty}C^{a}_{n}\,e^{in\varphi}\label{eqn25a1}
        \eea
    and in terms of the Fourier modes of the fields, the algebra of global
    charges now becomes:
        \bea
            \big\{B^{a}_{n},B^{b}_{m}\big\}_{D}&=&-f^{ab}_{c}B^{c}_{n+m}+
                            ing^{ab}\delta_{n+m}\nonumber\\
            \big\{B^{a}_{n},C^{b}_{m}\big\}_{D}&=&
                -f^{ab}_{c}C^{c}_{n+m}\label{eqn25a2}\\
            \big\{C^{a}_{n},C^{b}_{m}\big\}_{D}&=&0\nonumber
        \eea
    where $f^{ab}_{c}$ are the structure constants of the
    Poincar\'{e} algebra whose indices are raised and lowered with the the
    Cartan-Killing metric $g^{ab}=\tilde{Tr}(\gamma^{a}\gamma^{b})$.\\
\indent
    The first bracket in (\ref{eqn25a2}) can be recognized at once as the
    traditional central extension of the Ka\v{c}-Moody algebra of gauge
    charges of pure gravity. As it is now obvious, the central extension
    of the Ka\v{c}-Moody is further inhomogeneized by the generators
    $C^{a}_{n}$ that form an infinite dimensional abelian algebra, and whose
    brackets with the Ka\v{c}-Moody generators resemble (up to a sign) the
    brackets of the Poincar\'{e} translation generators with the generators
    of Lorentz rotations.\\
\indent
    Following now Dirac's quantization procedure, the quantum algebra of the
    operators $\hat{B}^{a}_{n}$, $\hat{C}^{a}_{n}$ is obtained by promoting the
    Fourier modes of the fields to operators and by defining the quantum
    commutators as $(-i)$ times the corresponding Dirac brackets. The resulting
    operator algebra is therefore given by the relations:
        \bea
            \big[\hat{B}^{a}_{n},\hat{B}^{b}_{m}\big]&=&if^{ab}_{c}\hat{B}^{c}_{n+m}+
                            ng^{ab}\delta_{n+m}\nonumber\\
            \big[\hat{B}^{a}_{n},\hat{C}^{b}_{m}\big]&=&
                -f^{ab}_{c}\hat{C}^{c}_{n+m}\label{eqn25a3}\\
            \big[\hat{C}^{a}_{n},\hat{C}^{b}_{m}\big]&=&0\nonumber
        \eea

\subsection{Field-dependent gauge parameters}

\indent
    We now consider the case of diffeomorphisms,and for simplicity reasons,
    we will investigate this case in the Poincar\'{e} formulation of the theory. It
    is a known fact that for topological field theories the diffeomorphism symmetries
    are equivalent on-shell to gauge symmetries with field-dependent gauge parameters.
    Under these circumstances, and since in the following we are only interested in
    the case of spatial diffeomorphisms, for this case the diffeomorphisms can be
    represented by gauge transformations whose gauge parameters depend on the fields
    of the theory through the following the relations:
        \bea
            \lambda^{a}&=& v\cdot {\cal{A}}^{a}=v^{A}{\cal{A}}^{a}_{A}\nonumber\\
            \gamma^{a}&=&v\cdot \Sigma^{a}=-v^{\varphi}\Sigma^{a}_{r\varphi}dr+
                        v^{r}\Sigma^{a}_{r\varphi}d\varphi\label{eqn26}
        \eea
    where in (\ref{eqn26}) $v$ is an arbitrary spatial vector, and we
    have used the notation $A=\{r,\varphi\}$ for the spatial
    indices of vectors and forms.\\
\indent
    Before proceeding with the calculations of the diffeomorphism charges, it
    is necessary to make a few useful remarks concerning the functional
    derivatives of the symmetry generators and their Poisson brackets for the
    case where the gauge parameters depend on the fields as described above.
    First of all, and referring to the calculations for the field-independent
    case, when calculating the first order variation of the symmetry generators
    $G_{JP}[\lambda]\equiv G_{JP}[v]$ and $G_{K}[\gamma]\equiv G_{K}[v]$, the
    field dependence of the gauge parameters will only introduce additional terms
    proportional to the constraints in the surface integrals, leaving all the
    boundary terms calculated earlier unchanged. Secondly, the very same thing
    happens when calculating the Poisson algebra of the differential symmetry
    generators. Hence, and since we are only interested in the Dirac algebra of
    global charges, we only have to worry about the processing of the respective
    boundary terms under the circumstances where the gauge parameters have the
    field dependence as described in (\ref{eqn26}). All the rest of the surface
    terms resulting from the Poisson algebra of the differentiable symmetry
    generators are proportional to constraints and therefore vanish identically
    on-shell.\\
\indent
    Repeating once again the analysis regarding the differentiability of the
    symmetry generators for the case of field-dependent parameters, it is easy
    to see that the differentiability of these generators can be ensured by
    adding to their first order variation the respective diffeomorphism
    infinitesimal charges:
        \bea
            \delta C_{JP}[v]&=&-\int_{\partial S}d\varphi
                    (v^{A}{\cal{A}}_{aA})\delta{\cal{A}}^{a}_{\varphi}\nonumber\\
            \delta C_{K}[v]&=&\int_{\partial S}d\varphi
                    (v\cdot\Sigma^{a})_{\varphi}\delta\Phi_{a}=
                    \int_{\partial S}d\varphi(v^{r}\Sigma^{a}_{r\varphi})\delta\Phi_{a}\label{eqn27}
        \eea
    where in (\ref{eqn27}) we have explicitly considered that the boundary
    $\partial S$ is a circle with $(r,\varphi)$ the radial and respectively angular
    coordinates, and we have used the notations $C_{JP}[v]$ and $C_{K}[v]$ for the
    diffeomorphism charges in order to distinguish them from the ones determined
    in the field independent case.\\
\indent
    Having found these infinitesimal charges only solves half of the differentiability
    problem of the symmetry generators, since in order to define such differentiable
    generators we must also determine the conditions under which the infinitesimal
    charges ar integrable. It is clear from (\ref{eqn27}) that these two infinitesimal
    charges are not trivially integrable anymore as in the field-independent case, and for
    this reason we need to address the issue of integrability of each of these charges
    separately.\\
\indent
    Consider the infinitesimal charge $\delta C_{JP}[v]$ in (\ref{eqn27}).
    By imposing the traditional $SL(2,R)$ boundary condition \cite{Banados1}:
        \be
            \delta{\cal{A}}^{a}_{r}=0\label{eqn28}
        \ee
    i.e. by fixing the radial components of the Poincar\'{e} connection on the
    boundary $\partial S$ (which works equally well for our present purposes),
    this infinitesimal charge can be integrated to yield:
        \be
            C_{JP}=-\int_{\partial S}d\varphi\Big[v^{r}{\cal{A}}_{ar}{\cal{A}}^{a}_{\varphi}+
                    \frac{1}{2}v^{\varphi}{\cal{A}}_{a\varphi}{\cal{A}}^{a}_{\varphi}\Big]+
                    C^{0}_{JP}\label{eqn29}
        \ee
    where $C^{0}_{JP}$ is a functional integration "constant" which will be specified
    at a later time. With the diffeomorphism charge (\ref{eqn29}), one can immediately
    define the differentiable diffeomorphism  symmetry generator:
        \be
            \tilde{G}_{JP}[v]=G_{JP}[v]+Q_{JP}[v]\label{eqn30}
        \ee
    and from this point on, the calculation of the Poisson algebra of this constraint
    with itself, and the corresponding Dirac algebra of the diffeomorphism charges is
    standard \cite{Banados1}, \cite{Park1}. A rather straightforward calculation
    yields for the Poisson bracket of this constraint with itself the expression:
        \be
            \Big\{\tilde{G}_{JP}[v],\tilde{G}_{JP}[w]\Big\}_{P.B.}=
                    \tilde{G}_{JP}\big[[v,w]^{A}\big]+\int_{\partial S}d\varphi
                    {\cal{A}}_{ar}{\cal{A}}^{a}_{r}v^{r}(\partial_{\varphi}w^{r})
                    \label{eqn31}
        \ee
    where in (\ref{eqn31}) we have used for the Lie bracket of
    vectors the notation $[v,w]^{A}=v^{B}\partial_{B}w^{A}-w^{B}\partial_{B}v^{A}$.\\
    Correspondingly, the Dirac algebra of global diffeomorphism
    charges will be given by the expression:
        \be
            \Big\{C_{JP}[v],C_{JP}[w]\Big\}_{D}=
                    C_{JP}\big[[v,w]^{A}\big]+\int_{\partial S}d\varphi
                    {\cal{A}}_{ar}{\cal{A}}^{a}_{r}v^{r}(\partial_{\varphi}w^{r})
                    \label{eqn32}
        \ee
    and this is, as expected, the traditional central extension of the Virasoro
    algebra of diffeomorphism charges of pure Poincar\'{e} gravity in Chern-Simons
    formulation. It should be noted that in obtaining (\ref{eqn31}), (\ref{eqn32})
    the integration ``constant" $C^{0}_{JP}$ has been chosen such that the
    boundary term in the r.h.s. of the brackets is independent of the (still unfixed)
    fields on the boundary. With this choice, the boundary term becomes the
    usual central charge of the Virasoro algebra.\\
\indent
    Consider now the infinitesimal charge $\delta C_{K}[v]$. A boundary
    condition, compatible with the classes of solutions for the $\Sigma\Phi EA$
    theory discussed in the previous section to require that the $\Phi^{A}$ fields
    be constant on the boundary $\partial S$, i.e. that:
        \be
            \delta\Phi^{a}=0\label{eqn33}
        \ee
    With this boundary condition, the infinitesimal charge can be trivially
    integrated to a functional "constant", and we have:
        \be
            C_{K}[v]=C_{K}^{0}\label{eqn34}
        \ee
    The functional integration "constant" need not be vanishing, and for the
    remainder of this section, we will assume that $C_{K}^{0}\neq 0$.\\
\indent
    Of course, we can formally define the differentiable symmetry generator:
        \be
            \tilde{G}_{K}[v]=G_{K}[v]+C_{K}^{0}\label{eqn35}
        \ee
    and we can proceed to calculate the Poisson brackets of this generator
    with itself and with the previous generator $G_{JP}[v]$.
    The Poisson bracket of $\tilde{G}_{K}[v]$ with itself is trivial, and
    can be read off directly from the corresponding bracket in (\ref{eqn24}):
        \be
            \big\{\tilde{G}_{K}[v],\tilde{G}_{K}[w]\big\}_{PB}=0\label{eqn36}
        \ee
    The Poisson bracket with the Poincar\'{e} generator $G_{JP}[v]$, this bracket
    can be easily be evaluated if we recall the observations made in the beginning
    of this subsection. According to these observations, the bracket we are
    interested in will contain a surface integral whose integrand is a linear
    combination of the constraints of the theory, plus the surface term of the
    corresponding field-independent case in which the constant gauge parameters
    are replace by the field dependent ones according to (\ref{eqn26}). Under
    these circumstances we can write:
        \be
            \big\{\tilde{G}_{JP}[v],\tilde{G}_{K}[w]\big\}_{PB}=
                    \int_{S}d^{2}x[\sim constraints]+
                    \int_{\partial S}d\varphi
                    f_{abc}(v^{A}{\cal{A}}^{a}_{A})(w^{r}\Sigma^{b}_{r\varphi})\Phi^{c}
                    \label{eqn37}
        \ee
\indent
    Unfortunately, and in contrast to the field-independent case, the Poisson bracket
    in (\ref{eqn37}) cannot be put in a nice closed form that exhibits explicitly
    the structure of the algebra. For this reason, we will ignore the Poisson algebra
    of the differentiable diffeomorphism constraints, and will focus on the main goal
    of this subsection which is the determination of the Dirac algebra of global
    diffeomorphism charges.\\
\indent
    To this end, it is easy to show that on-shell, due to the equations of motion for
    the $\Phi^{a}$ fields on the boundary (where these fields are constant), the
    boundary term in (\ref{eqn37}) vanishes identically. Under these circumstances,
    it follows from (\ref{eqn32}), (\ref{eqn36}) and (\ref{eqn37}) that the Dirac
    algebra of the diffeomorphism charges is formally given by:
        \bea
            \Big\{C_{JP}[v],C_{JP}[w]\Big\}_{D}&=&
                    C_{JP}\big[[v,w]^{A}\big]+\int_{\partial S}d\varphi
                    {\cal{A}}_{ar}{\cal{A}}^{a}_{r}v^{r}(\partial_{\varphi}w^{r})\nonumber\\
            \Big\{C_{JP}[v],C_{K}[w]\Big\}_{D}&=&0\label{eqn38}\\
            \Big\{C_{K}[v],C_{K}[w]\Big\}_{D}&=&0\nonumber
        \eea

\indent
    As in the case of field-independent gauge parameters, it is traditional to
    rewrite the algebra of diffeomorphism charges (\ref{eqn38}) in terms of the
    Fourier modes of the fields that are free on the boundary. However, before
    proceeding with any further considerations, it is useful to note that since in
    (\ref{eqn38}) the algebra of the Poincar\'{e} charges $C_{JP}$ is trivially
    inhomogeneized by an abelian (constant) charge $C_{K}$, we only have to worry
    about the Fourier modes of the Poincar\'{e} diffeomorphism charges. This is a
    very convenient situation indeed, because this issue has been extensively
    studied in the literature. For these reasons, we will only quote the results
    that are relevant to our discussion, referring the interested reader for details
    to \cite{Banados1}.\\
\indent
    The Fourier modes $L_{n}$ of the Poincar\'{e} charges can be obtained from the
    Fourier expansion of these charges, and they can be shown to satisfy the Dirac
    bracket:
        \be
            \big\{L_{n},L_{m}\big\}_{D}=i(n-m)L_{n+m}+
                i({\cal{A}}_{ar}{\cal{A}}^{a}_{r})n(n^{2}-1)\delta_{n+m}\label{eqn38a1}
        \ee
    where it must be kept in mind that ${\cal{A}}^{a}_{r}=\alpha^{a}$ is a constant on the
    boundary, and therefore $\alpha^{2}={\cal{A}}_{ar}{\cal{A}}^{a}_{r}$ plays the
    role of a classical algebraic charge. In the form (\ref{eqn38a1}), the algebra
    of the Fourier modes can be recognized as the central extension of the classical
    Virasoro algebra, with the charge given by $\alpha^{2}$.\\
\indent
    The quantization of the Virasoro algebra is more involved than the quantization
    of the previous Ka\v{c}-Moody algebra, due to operator ordering problems, and for
    this reason it requires more detailed consideration.\\
\indent
    The problems in quantizing the Virasoro algebra arise from the
    fact that the Virasoro generators $L_{n}$ are quadratic in the
    generators of the Ka\v{c}-Moody algebra. Indeed, this can easily
    be seen from the expression of the Poincar\'{e} charge in (\ref{eqn29})
    if we introduce the Fourier expansion (\ref{eqn25a1}) for the
    boundary connection. Explicitly, we obtain \cite{Banados1}:
        \be
            L_{n}=\frac{1}{2}\sum_{m}B_{am}B_{n-m}^{a}+in\alpha_{a}B^{a}_{n}+
                    \frac{1}{2}\alpha^{2}\delta_{n}\label{eqn38b1}
        \ee
    and it is clear from (\ref{eqn38b1}) that if we were to construct the
    operator version of $L_{n}$ by directly replacing the $B^{a}_{n}$
    generators with the corresponding quantum operators operators, we
    could run into potential singularity issues due to the fact
    that both Ka\v{c}-Moody in the quadratic term are evaluated at
    the same point on the boundary.\\
\indent
    The solution to these singularity issues is to use the
    Sugawara construction. According to this construction
    \cite{Goddard1}, one needs to introduce a normal ordering for
    the operators corresponding to the Ka\v{c}-Moody algebra
    generators - traditionally the ordering requires that the operators
    with positive indices $m$ to be on the right - that will regularize the
    infinities. However, by simply introducing a normal ordering
    for the operators associated with the Ka\v{c}-Moody generators
    solves only half of the issue of quantizing the Virasoro
    algebra, for the simple reason that the normal ordered operators
    $(:L_{n}:)$ obtained through this procedure do not obey the
    commutation relations of a Virasoro algebra anymore.\\
\indent
    Nevertheless, we can solve this last issue by defining the operators:
        \be
            \hat{L}_{n}=\tilde{\beta}:L_{n}:+\tilde{a}\delta_{n}\label{eqn38a2}
        \ee
    where $(:\ :)$ stands for normal ordering, $\tilde{\beta}=[1+\frac{1}{2}Q_{2}]^{-1}$,
    $\tilde{a}=\frac{1}{2}\alpha^{2}\tilde{\beta}(\tilde{\beta}-1)$, and $Q_{2}$ is
    the quadratic Casimir invariant of the Poincar\'{e} algebra in the adjoint
    representation. The newly defined operators $\hat{L}_{n}$ now satisfy the quantum
    Virasoro algebra:
        \be
            \big[\hat{L}_{n},\hat{L}_{m}\big]=(n-m)\hat{L}_{n+m}+
                qn(n^{2}-1)\delta_{n+m}\label{eqn38a3}
        \ee
    where now $q$ is a quantum central charge which is different from the classical
    central charge $\alpha^{2}$. For the theory of gravity under present consideration
    \footnote{Of particular importance in the calculation of the quantum charge is the
    dimension of the Lie algebra underlying the theory of gravity under
    consideration - in this case the Poincare algebra. For more details about the
    general dependence of the Virasoro central charge on the dimension of the
    underlying Lie algebra, the reader is referred to \cite{Banados1}, \cite{Goddard1}.},
    it can be shown that the quantum central charge of the quantum Virasoro algebra
    (\ref{eqn38a3}) is in fact given by the  expression:
        \be
            q=\alpha^{2}\beta^{2}+\frac{\beta}{2}\label{eqn38b2}
        \ee
\indent
    From (\ref{eqn38b1}) it can be seen the quantum global charge
    contains two terms. The first term which is nothing else than the
    classical central charge rescaled by the square of the "renormalization"
    factor $\beta$ that has been introduced in order to define the
    operators associated with the Virasoro generators in
    (\ref{eqn38a2}). The second term in the expression of the
    quantum central charge is the direct consequence of the
    Sugawara construction, and as such it has an entirely quantum
    character. It should be noted at this time that due to the
    fact that the classical Virasoro algebra is trivially
    inhomogeneized by the abelian algebra of the charhes
    associated with the topological matter fields, it comes at no
    surprise the fact that the matter fields of the $\Sigma\Phi
    EA$ theory have no influence upon the quantum central charge
    of gravity. This is the direct consequence of the boundary
    conditions that have been chosen in order to determine the
    diffeomorphism charges.\\
\indent
    Having the quantum Virasoro algebra, it is trivial to obtain
    the quantum algebra of the Fourier modes corresponding to the
    diffeomorphism charges in the $\Sigma\Phi EA$ model. it is
    given by the relations:
        \bea
            \big[\hat{L}_{n},\hat{L}_{m}\big]&=&(n-m)\hat{L}_{n+m}+
                qn(n^{2}-1)\delta_{n+m}\nonumber\\
            \big[\hat{L}_{n},\hat{C}_{K}\big]&=&0\label{eqn38a4}\\
            \big[\hat{C}_{K},\hat{C}_{K}\big]&=&0\nonumber
        \eea

\section{The $\Sigma\Phi EA$ Model as a generalized Chern-Simons theory}

\indent
   As noted earlier, it has thus far proven impossible to formulate
   the $\Sigma\Phi EA$ theory as either a BF theory or
    as a traditional Chern-Simons theory. However, as we will prove in the following,
    the theory can be formulated as a generalized Chern-Simons theory with a multiform
    connection involving both bosonic and fermionic matter fields, defined over
    an algebra that is not the algebra of the constraints.\\
\indent
    In order to understand how such a particular formulation arises naturally for the
    $\Sigma\Phi EA$ theory, it is worth to begin by illustrating the difficulties that
    one faces in reformulation of the theory as a Chern-Simons theory.\\
\indent
    The first issue that one must deal with in attempting such a formulation is the form
    content of the matter fields. Since the matter fields are 0-forms and 2-forms
    respectively, any generalized connection defined using these fields will necessarily
    be a multiform connection. This presents a major problem, since such a multiform
    connection requires generally the introduction of additional de Rham currents
    in order to be able to define a generalized holonomy over some submanifold of the
    spacetime manifold, submanifold which usually is not a closed loop, as in the standard
    Chern-Simons theory.\\
\indent
    Furthermore, even if one ignores the above problems, and defines such a multiform
    generalized connection, for the particular case of the $\Sigma\Phi EA$, if one
    attempts to define this generalized connection over the Lie algebra generated by the
    Poisson brackets of the constraints, it is rather obvious that the action of the
    $\Sigma\Phi EA$ cannot be actually written as a Chern-Simons action. This is most
    easily seen from the following argument. Assume that we define a generalized
    connection form:
        \be
            {\cal{A}}=A^{i}\bar{J}_{i}+E^{i}\bar{P}_{i}+
                    \Phi^{i}\bar{Q}_{i}+\Sigma^{i}\bar{R}_{i}\label{tcsconn}
        \ee
    where $(\bar{J}_{i},\bar{P}_{i},\bar{Q}_{i},\bar{R}_{i})$  are the generators of
    the constraint algebra of the $\Sigma\Phi EA$ model (\ref{sfeaalg}), on which we
    introduce the invariant non-degenerate bilinear form:
        \bea
            \tilde{Tr}(\bar{J}^{i}\bar{P}^{j})=\eta^{ij},\;
            \tilde{Tr}(\bar{Q}^{i}\bar{R}^{j})=\eta^{ij},\label{scalprod}
        \eea
    with all the rest of the pairings vanishing.  When calculating the
    derivative term $\frac{1}{2}\;{\cal{A}}\wedge d{\cal{A}}$ in the Chern-Simons
    action, it will contain explicitly the terms
    $\frac{1}{2}\;(\Phi^{i}\wedge d\Sigma_{i}+\Sigma^{i}\wedge d\Phi_{i})$, and
    by using integration by parts these terms combined should yield the
    term $\Sigma^{i}\wedge d\Phi_{i}$ as the first component of
    the covariant derivative of the topological matter fields in
    the action and an additional surface term. It is clear however
    that due to the form content of the above terms involving
    the topological matter fields, the integration by parts will
    only yield a surface term since the two resulting terms
    involving the exterior derivatives of the fields $\Sigma$,
    $\Phi$ will cancel each other. Furthermore, it is also clear
    from the above argument that in order to be able to write the
    $\Sigma\Phi EA$ model as a Chern-Simons theory, i.e. in order to recover
    the component $\Sigma^{i}\wedge d\Phi_{i}$ from the "derivative"
    Chern-Simons term $\frac{1}{2}\;{\cal{A}}\wedge d{\cal{A}}$, one must
    use a formalism which combines either fermionic matter fields with a
    "regular" Lie algebra, or alternatively, bosonic matter fields with a
    graded Lie algebra.\\
\indent
    Fortunately, such a formalism that generalizes the
    Chern-Simons theory to include both bosonic and fermionic
    fields with a graded gauge Lie algebra has been developed
    \cite{KaWa}. Using this formalism, we will show
    that the $\Sigma\Phi EA$ model can be written
    as such a generalized Chern Simons theory if the topological
    matter fields are considered to be of the fermionic type.

\subsection{The generalized Chern-Simons formalism}

\indent
    We begin by briefly reviewing the generalized Chern-Simons
    formalism developed by Kawamoto and Watabiki \cite{KaWa},
    whose original purpose was to both extend the Chern-Simons
    formalism to higher-dimensional spacetime manifolds, and at
    the same time to extend it to higher order tensorial
    connections, even in (2+1)-dimensional spacetimes.\\
\indent
    One starts with a generalized connection form $\cal{A}$ and a
    generalized gauge parameter $\nu$ that include both bosonic and
    fermionic type of fields, which are defined as follows:
        \bea
            {\cal{A}}&=&{\textbf{1}}F+{\textbf{i}}\tilde{F}+
                            {\textbf{j}}B+{\textbf{k}}\tilde{B}\nonumber\\
                 \nu &=&{\textbf{1}}\tilde{b}+{\textbf{i}}b+
                            {\textbf{j}}\tilde{f}+{\textbf{k}}f\label{kawacon}
        \eea
    where $f,F$ are fermionic odd-rank form fields,
    $\tilde{f},\tilde{F}$ are fermionic even-rank form fields,
    $b,B$ are odd-rank bosonic fields, and $\tilde{b},\tilde{B}$ are
    bosonic even-form fields. This means for example that the
    bosonic field $B$ can be written formally as
    $B=\sum_{p-odd}B_{(p)}$ where $B_{(p)}$ are p-rank bosonic forms
    with p odd. Of course, similar such formal relations can be
    written for each of the fields in (\ref{kawacon}), with p
    being, as the definition of the fields dictates, odd or even
    numbers. The symbols $(\textbf{1},\textbf{i},\textbf{j},\textbf{k})$
    in (\ref{kawacon}) are the "generators" of the quaternionic
    "generalized algebra" defined as:
        \bea
            {\textbf{1}}^{2}=\textbf{1},\;
            {\textbf{i}}^{2}=\varepsilon_{1}\textbf{1},\;
            {\textbf{j}}^{2}=\varepsilon_{2}\textbf{1},\;
            {\textbf{k}}^{2}=-\varepsilon_{1}\varepsilon_{2}\textbf{1},\nonumber\\
            {\textbf{i}}{\textbf{j}}=-{\textbf{j}}{\textbf{i}}=\textbf{k},\;
            {\textbf{j}}{\textbf{k}}=-{\textbf{k}}{\textbf{j}}=-\varepsilon_{2}\textbf{k},\;
            {\textbf{k}}{\textbf{i}}=-{\textbf{i}}{\textbf{k}}=-\varepsilon_{1}\textbf{j}.\label{gqalg}
        \eea
    and the coefficients $(\varepsilon_{1},\varepsilon_{2})$ can
    take the values $(-1,-1)$ ,in which case the algebra defined
    by these generators becomes the traditional quaternion
    algebra), or $(-1,1)$, $(1,-1)$, $(1,1)$ in which case the
    algebra becomes the $gl(2,R)$ Lie algebra.\\
\indent
    One also introduces a graded gauge Lie algebra, with commuting and
    anticommuting generators $(T_{m})$ and $(S_{\mu})$ respectively,
    defined as:
        \bea
            \{T_{m}, T_{n}\}_{-}&=&{c_{mn}}^{p} T_{p}\nonumber\\
            \{T_{m}, S_{\mu}\}_{-}&=&{g_{m\mu}}^{\nu} S_{\nu}\label{gradalg}\\
            \{S_{\mu}, T_{\nu}\}_{+}&=&{h_{\mu\nu}}^{p} T_{p}\nonumber
        \eea
    where the $\pm$ indices at the right of the Poisson brackets
    in (\ref{gradalg}) indicate the commuting and anticommuting
    character of the brackets involved. The structure
    constants obey the corresponding graded Jacobi identities.
    To simplify the notation, in the following we will drop the
    exterior (wedge) product symbol from the mathematical relations,
    its existence being implicitly assumed everywhere where multiplication of
    forms is involved. Also, in order to keep the consistency with the index
    notations used in the previous sections, at this time
    we introduce the following conventions. All Latin
    lower case indices from the end of the alphabet $(m,n,p,...)$
    and all Greek lower case indices are now Lie algebra formal
    indices, and we will use Latin indices and Greek indices to
    differentiate between the commuting and anticommuting algebra
    generators. All sums involving such indices are purely formal
    in this context, and do not reflect the explicit structure of
    the gauge Lie algebra and fields involved in the formalism.
    Later on, when the graded gauge Lie algebra and field structure
    for the $\Sigma\Phi EA$ theory are introduced, all the formal
    expressions will be made explicit by returning to the previous
    index convention with only latin lower case indices
    $(i,j,k,...)$ as Lie algebra indices.\\
\indent
    With the graded gauge algebra (\ref{gradalg}), one introduces
    the following internal structure for the fermionic and bosonic
    fields involved in the definition (\ref{kawacon}) of the generalized
    connection and gauge parameter:
        \bea
            F=F^{\mu}S_{\mu},\; \tilde{F}=\tilde{F}^{m}T_{m},\;
            B=B^{m}T_{m},\; \tilde{B}=\tilde{B}^{\mu}S_{\mu},\nonumber\\
            f=f^{m}T_{m},\; \tilde{f}=\tilde{f}^{\mu}S_{\mu},\;
            b=b^{\mu}S_{\mu},\; \tilde{b}=\tilde{b}^{m}T_{m}.\label{kawafields}
        \eea
    and it should be noted at this time that the model takes into
    consideration all possible combinations of fields and algebra
    generators, i.e. bosonic and fermionic fields with
    commuting/bosonic algebra generators and bosonic and
    fermionic fields with anticommuting/fermionic algebra
    generators.\\
\indent
    With the structure introduced above, we can now define the
    generalized Chern-Simons action:
        \be
            S_{gen}=\int_{M}Tr^{*}[\frac{1}{2}{\cal{A}}Q({\cal{A}})+
                                    \frac{1}{3}{\cal{A}}^{3}]\label{kawaCS}
        \ee
    where $M$ is a spacetime manifold having an arbitrary finite
    dimension, and $Q$ is a nilpotent generalized derivative operator
    given by the expression:
        \be
            Q={\textbf{j}}d \label{kawader}
        \ee
    with $d$ the traditional exterior derivative.\\
\indent
    The invariant non-degenerate bilinear form $Tr^{*}$ that
    appears in the definition of the generalized Chern-Simons action
    (\ref{kawaCS}) is defined as follows. One first introduces an
    invariant and non-degenerate (traditional) bilinear form $Tr$ on
    the graded algebra (\ref{gradalg}), and once and if such a bilinear
    form has been introduced, then the extended bilinear form
    $Tr^{*}$ is defined as the projection of the terms in the
    integrand (with the appropriate dimensionality in accordance to
    the dimension of the spacetime manifold) on one of the
    generators of the quaternion algebra. For example, if one
    chooses to use in the trace the projection along $\textbf{i}$,
    then one has:
        \be
            Tr^{*}({\cal{A}})\equiv Tr_{\textbf{i}}({\cal{A}})=
                    Pr_{\textbf{i}}({\cal{A}})=\tilde{F}\label{kawatr1}
        \ee
    and it is clear from these considerations that with the above
    definition for the generalized invariant bilinear form, one
    can in fact have four different such bilinear forms, each
    corresponding to one of the generators of the generalized
    quaternionic algebra (\ref{gqalg}).\\
\indent
    Furthermore, such a generalized bilinear form must also obey
    certain constraints, in order for resulting Chern-Simons formalism
    to be internally consistent. The principal constraint that
    must be imposed on the generalized bilinear form derives from
    the requirement that when calculating the explicit form of the cubic
    term in the generalized Chern-Simons action, the generalized
    connection should obey the consistency condition
    ${\cal{A}}^{2}{\cal{A}}={\cal{A}}{\cal{A}}^{2}$. A long but
    straightforward calculation yields the following conditions
    that must be obeyed by the generalized bilinear form:
        \bea
            Tr_{\textbf{1}}(\{T_{m},S_{\mu}\}_{-})=
                    Tr_{\textbf{k}}(\{T_{m},S_{\mu}\}_{-})=0 \nonumber\\
            Tr_{\textbf{i}}(\{T_{m},T_{n}\}_{-})=
                    Tr_{\textbf{i}}(\{S_{\mu},S_{\nu}\}_{+})=0 \label{kawatr2}\\
            Tr_{\textbf{j}}(\{T_{m},T_{n}\}_{-})=
                    Tr_{\textbf{j}}(\{S_{\mu},S_{\nu}\}_{+})=0 \nonumber
        \eea
    and from (\ref{kawatr2}) it is obvious that in fact these
    consistency conditions on the generalized trace translate in
    conditions that must be imposed at the level of the
    traditional non-degenerate invariant bilinear form defined on
    the graded Lie algebra\footnote[2]{In the original work of Kawamoto
    and Watabiki \cite{KaWa}, they also require consistency
    conditions similar to (\ref{kawatr2}) for higher order
    products of generators of the graded gauge algebra, which are
    necessary for the gauge invariance of the generalized action.
    However, in redoing the calculations, we have found no need to
    introduce such higher order trace conditions.}. Following the notation in \cite{KaWa},
    this means that we can define two types of bilinear forms on
    the underlying graded gauge Lie algebra (\ref{gradalg}). For
    projections along the quaternionic generators
    $\textbf{i}$ and $\textbf{j}$ the trace on the graded gauge
    algebra is denoted by $STr$ (supersymmetric trace by analogy with traditional
    supersymmetric theories), while for projections along the
    quaternionic generators $\textbf{1}$ and $\textbf{k}$ the
    trace on the graded gauge algebra is denoted by $HTr$
    (heterotic trace).\\
\indent
    Under these circumstances, and no matter which component of
    the quaternion algebra we choose in defining the generalized
    Chern-Simons action, the equations of motion of (\ref{kawaCS})
    are given by the vanishing of the curvature of the generalized
    connection form:
        \be
            {\cal{F({\cal{A}})}}\equiv
                    Q({\cal{A}})+{\cal{A}}^{2}=0 \label{kawaeom}
        \ee
    and the action is invariant under the generalized gauge
    transformations:
        \be
            \delta {\cal{A}}=Q(\nu)+\{{\cal{A}},\nu\}_{-}\label{kawaCSgge}
        \ee
\indent
    As mentioned above, one can define four generalized traces for the
    action (\ref{kawaCS}), depending on which generator of the
    quaternion algebra is chosen for projection. It is not
    difficult to see that in this way each of the four resulting
    actions has a definite dimension type, i.e. corresponds to
    the action on an odd or even dimensional manifold, and a
    definite fermionic or bosonic character. For example, if one
    defines the generalized trace as $Tr^{*}(...)\equiv
    Str_{\textbf{j}}(...)$, following again the syntax in
    \cite{KaWa}, one obtains a bosonic action defined on an
    odd-dimensional spacetime manifold M. Of course, choosing
    other quaternionic generator with the appropriate type of trace
    on the graded gauge algebra, one can also obtain bosonic
    actions defined on even dimensional spacetime manifolds, and
    fermionic action defined on even and odd dimensional
    manifolds, but such actions will not be considered here.\\
\indent
    Since the $\Sigma\Phi EA$ theory is a (2+1)-dimensional
    theory, and since its action is manifestly bosonic, we are
    only interested in the generalized bosonic Chern-Simons action
    defined on a (2+1)-dimensional manifold defined as above.
    Choosing the field content of the generalized connection such
    that it contains only bosonic odd-rank forms and fermionic
    even-rank forms (i.e. $F=\tilde{B}=0$), the formal expression of
    this action in terms of the generators of the graded gauge
    algebra is given by:
        \be
            S_{bo}=\int_{M}Str_{\textbf{j}}[\frac{1}{2}{\cal{A}}Q({\cal{A}})+
                                    \frac{1}{3}{\cal{A}}^{3}]=
                    \int_{M}Str[{\cal{L}}_{\textbf{j}}]\label{kawabodd1}\\
        \ee
    where the argument of the trace has the expression
        \bea
            {\cal{L}}_{\textbf{j}}&=&\{\varepsilon_{2}[\frac{1}{2}\,B^{m}dB^{n}+
                    \frac{1}{6}\,{c_{pq}}^{m}B^{p}B^{q}B^{n}](T_{m}T_{n})-\varepsilon_{1}[
                    \frac{1}{2}\,{\tilde{F}}^{m}d{\tilde{F}}^{n}+{}\nonumber\\
                    &&{}+\frac{1}{3}\,({c_{pq}}^{m}{\tilde{F}}^{p}B^{q}{\tilde{F}}^{n}-
                    \frac{1}{2}\,{c_{pq}}^{m}{\tilde{F}}^{p}{\tilde{F}}^{q}B^{n})](T_{m}T_{n})\}
                    \label{kawabodd2}
        \eea
\indent
    Having established the relations (\ref{kawabodd1}) and (\ref{kawabodd2})
    we can now proceed with the proof that the $\Sigma\Phi EA$ model can
    be written as such a generalized Chern-Simons theory with
    fermionic topological matter fields.

\subsection{The $\Sigma\Phi EA$ model in the context of the
generalized Chern-Simons formalism}

\indent
    The relations (\ref{kawabodd1}) and (\ref{kawabodd2})
    established in the previous section are as far as one can go,
    within the general framework of the extended formalism
    developed in \cite{KaWa}, in proving that the $\Sigma\Phi EA$
    model can be written as such a Chern-Simons theory. In order
    to make any further progress it becomes absolutely
    necessary to introduce the explicit forms of the graded gauge
    algebra and of the fields appearing in the generalized connection
    (\ref{kawacon}).\\
\indent
    As far as the field content of the generalized connection is
    concerned, this issue is quite straightforward. The fields
    that we have used in obtaining the expression
    (\ref{kawabodd2}) of the odd-dimensional bosonic action are
    the odd-rank bosonic field $B$ and the even-rank fermionic
    field $\tilde{F}$ and by simple comparison with the original
    $\Sigma\Phi EA$ action (\ref{ssfea}), it is clear that the
    internal structure of these fields can only be of the form:
        \bea
            B&=&A^{i}\bar{J}_{i}+E^{i}\bar{P}_{i}\equiv B^{m}T_{m}\nonumber\\
            \tilde{F}&=&\Phi^{i}\bar{Q}_{i}+\Sigma^{i}\bar{R}_{i}\equiv {\tilde{F}}^{m}T_{m}
                            \label{kawaflds}
        \eea
    where $\{\bar{J}_{i},\bar{P}_{i}\}$ are the generators of the Poincar\'{e}
    algebra, and $\{\bar{Q}_{i},\bar{R}_{i}\}$ are two additional sets of
    (still commuting/bosonic) generators of the gauge Lie algebra,
    whose commutation relations have yet to be specified.\\
\indent
    After specifying the fields, we are left with the much mode
    difficult task of specifying the underlying gauge algebra.
    Of course, the only way to determine the ``correct" algebra is
    by trial and error, so what we have to do is to look for an
    algebra which when inserted in (\ref{kawabodd2}) yields as a
    final result the original $\Sigma\Phi EA$ action
    (\ref{ssfea}). By simple examination of (\ref{kawabodd2}) and
    (\ref{kawaflds}), it is clear that the underlying gauge
    algebra is not a graded algebra, but a regular Lie algebra,
    and under these circumstances the generalized trace $STr$
    reduces to the usual trace on a Lie algebra. Furthermore,
    it is also clear that this gauge algebra has to
    be an extension of sorts of the Poincar\'{e} algebra, and for
    this reason it is only natural that we should first check the
    constraint algebra (\ref{sfeaalg}) of the original theory.
    Unfortunately, it is not very difficult to show that with the
    constraint algebra (\ref{sfeaalg}) on which we have introduced
    the non-degenerate invariant bilinear form (\ref{scalprod}),
    the action (\ref{kawabodd1}) does not yield the action
    (\ref{ssfea}) of the original $\Sigma\Phi EA$ theory.\\
\indent
    However, if instead of the constraint algebra of
    the $\Sigma\Phi EA$ theory we use the constraint algebra (\ref{alg}) of
    the BCEA model with its non-degenerate invariant bilinear form
    (\ref{scalprod}), the situation changes. Using this algebra,
    and restoring the exterior product symbol and our original index
    convention where latin lower case indices $(i,j,k,...)$ are explicit Lie
    algebra indices of the dimensionally correct terms in
    (\ref{kawabodd2}), the odd bosonic action (\ref{kawabodd1})
    with the trace defined by (\ref{scalprod}) reduces to:
        \bea
            S_{bo}=\int_{M}\{\varepsilon_{2}(E_{i}\wedge R^{i}[A]+
                    \frac{1}{2}\,d[E_{i}\wedge A^{i}])-\varepsilon_{1}(\Sigma_{i}\wedge D\Phi^{i}+
                    \frac{1}{2}\,d[\Sigma_{i}\wedge \Phi^{i}])\}\label{kawabodd3}
        \eea
    and it is clear that by setting $(\varepsilon_{1},\varepsilon_{2})=(-1,1)$
    in the quaternionic algebra, the action (\ref{kawabodd3})
    becomes identical (up to surface terms) to the original action (\ref{ssfea}) of the
    $\Sigma\Phi EA$ theory.

\section{Discussion and conclusions}

\indent
    In this paper we have considered a model (the $\Sigma\Phi EA$ model) of scalar and
    tensorial topological matter - represented by 0-form and 2-form fields - coupled
    minimally to gravity in (2+1) dimensions , and we have investigated its classical
    structure while at the same time comparing it, whenever possible, with a similar
    model (the BCEA model), involving only 1-form matter fields, that has already been
    studied in the literature. We have shown that the $\Sigma\Phi EA$ model has non-trivial
    classical sectors, in which the dynamics of the matter fields cannot be decoupled
    from gravity, and we have illustrated these sectors with two geometries, one
    corresponding to the BTZ black-hole, and the other one corresponding to FRW
    homogeneously/inhomogeneously expanding cosmological geometries.\\
\indent
    For the case of the BTZ geometry, we have calculated the Noether charges associated
    with the asymptotic symmetries, and have shown that these charges exhibit similar
    characteristics to the corresponding charges in the BCEA theory. Explicitly,
    in the case of the BCEA model, the mass and the angular momentum of the singularity
    exchange roles in the expressions of the Noether charges, such that the mass
    parameterizes the conserved angular momentum charge, and the angular momentum
    parameterizes the conserved energy. One strange implication of this role change
    of mass and angular momentum is that under certain conditions, the asymptotic mass
    can become smaller than the mass of the singularity, as if the matter fields were
    ``screening" the mass of the singularity. In the case of the $\Sigma\Phi EA$ model,
    the effect is even more drastic. The conserved charges both vanish, and while this
    may not be so strange for the angular momentum charge, it is definitely strange
    for the case of the  mass charge. In this case, the mass is completely obscured by
    the matter fields, to the point where the singularity simply disappears for any
    asymptotic observer.\\
\indent
    At the present time we have no underlying explanation for this mass ``screening"
    effect. It is, however, extremely interesting  that a similar effect
    appears to  exist in (3+1)-dimensional gravity when one considers its topological
    aspects \cite{Starodub1}, \cite{Starodub2}. However, the implications of this apparent
    similarity for gravity in (3+1) dimensions are not yet known and will
    require further investigation.\\
\indent
    For the case of homogeneously/inhomogeneously expanding FRW geometries, it would
    appear that the $\Sigma\Phi EA$ theory is the first theory that admits
    such solutions in the presence of matter, and as such it would be interesting to
    pursue this aspect further and in more detail.\\
\indent
    While the full issue of quantization of the $\Sigma\Phi EA$ theory has been deferred
    to a companion paper \cite{Mann2}, we have also studied, as a prelude to the full
    quantization of the theory, the global gauge charges associated with the constraints.
    Our analysis has shown that the classical algebras of charges are inhomogeneizations
    of the corresponding Ka\v{c}-Moody and Virasoro algebras of pure gravity. Furthermore,
    we have quantized the resulting charge algebras, and have shown that with the boundary
    conditions we have chosen, the quantum charge associated with the Virasoro subalgebra
    of the diffeomorphism algebra in the $\Sigma\Phi EA$ theory is identical to the quantum
    central charge of pure gravity.\\
\indent
    Finally, we have shown that while the $\Sigma\Phi EA$ theory cannot be formulated
    as either a traditional BF or a Chern-Simons theory - in contrast to the
    BCEA theory - it is still possible to formulate it as a generalized Chern-Simons
    theory with a multiform connection containing both bosonic and fermionic matter
    fields defined over the Lie algebra of I[ISO(2,1)] - which is different from the
    constraint algebra - with the help of the generalized quaternion algebra as an
    auxiliary algebra. While this formulation offers an extension of the theory to
    fermionic fields, more detailed investigation is necessary in order to fully
    understand its implications. One implication of this formulation, as it
    is apparent from our analysis if we follow the approach in \cite{Izqu1} is
    that the classical BTZ geometry could be described at the quantum level by a
    combination of fermionic and bosonic matter fields. Another implication, which
    is more far more reaching in its consequences is that this formulation could
    offer the possibility for the generalization of the concept of holonomy to a
    (multiform) connection and to higher dimensional submanifolds of the spacetime
    manifold.\\

Acknowledgements\\

\noindent
    The authors would like to thank L. Freidel for many useful discussions during
    the work on this paper, and also to A. Starodubtsev for discussions on the
    topological character of (3+1)-dimensional gravity. This work was supported in
    part by the Natural Sciences and Engineering Research Council of Canada.


\begin{thebibliography}{99}

\bibitem{Witten1} E.\ Witten, Nucl. Phys. {\bf B 311}, 46 (1988-1989).

\bibitem{Horowitz} Gary T.\ Horowitz, Commun. Math. Phys {\textbf
125}, 417-437 (1989).

\bibitem{bcea}  S.\ Carlip, J.\ Gegenberg, Phys. Rev.
{\bf D 44}, 424 (1991).

\bibitem{bceabtz}  S.\ Carlip, J.\ Gegenberg, R.\ B.\ Mann, Phys.
Rev. {\bf D 51}, 6854 (1995).

\bibitem{GKL} J.\ Gegenberg, G.\ Kunstatter, H.\ P.\ Leivo,
Phys. Lett. {\bf B 252(3)}, 381 (1990).

\bibitem{Teit} C.\ Teitelboim, M.\ Henneaux, Quantization of gauge
systems, Princeton University Press, Princeton, New Jersey,
(1992).

\bibitem{Henn} M.\ Henneaux, C.\ Teitelboim, J.\ Zanelli,
Nucl.Phys. {\bf B332}, 169, (1990).

\bibitem{Freidel1} L.\ Freidel, R.\ B.\ Mann, E.\ M.\ Popescu,
Clas.Quant.Grav. \textbf{22}, 3363, (2005).

\bibitem{btz1} M.\ Banados, C.\ Teitlboim, J.\ Zanelli
Phys. Rev. Lett. {\bf 69}, 1849 (1992); M.\ Banados, M.\ Henneaux
C.\ Teitlboim, J.\ Zanelli Phys. Rev. {\bf D 48}, 1506 (1993).

\bibitem{btz2}D. Cangemi, M. Leblanc and R.B. Mann, Phys. Rev.
{\bf D 48}, 3606 (1993).

\bibitem{Wald1} R.\ M.\ Wald, Phys. Rev. {\bf D 48}, 3427 (1993).

\bibitem{Wald2} J.\ Lee, R.\ M.\ Wald, J. Math. Phys. {\bf 31}, 725 (1990).

\bibitem{Wald3} V.\ Iyer, R.\ M.\ Wald, Phys. Rev. {\bf D 50}, 846 (1993).

\bibitem{Banados1} M.\ Banados, Phys. Rev. \textbf{D52}, 5816, (1996).

\bibitem{Park1} Mu-In Park, Nucl. Phys. \textbf{544}, 377, (1999).

\bibitem{Goddard1} P.\ Goddard, D. Olive, "Ka\v{c}-Moody and
Virasoro algebras: A reprint volume for physicists.", Adv. Ser.
Math. Phys. \textbf{3}, World Scientific Publishing Co., (1988).

\bibitem{KaWa} N.\ Kawamoto, Y.\ Watabiki Mod. Phys. Lett. \textbf{A7}, 1137, (1992).

\bibitem{Birmingham} D.\ Birmingham, M.\ Blau, R.\ Rakowski, G.\ Thompson,
Phys. Rep. \textbf{209}, 129, (1991).

\bibitem{Starodub1} A.\ Starobubtsev, preprint hep-th/0501191.

\bibitem{Starodub2} A.\ Starodubtsev, private communication.

\bibitem{Mann2} R.\ B.\ Mann, E.\ M.\ Popescu in preparation.

\bibitem{Izqu1} J.\ M.\ Izquierdo, P.\ K.\ Townsend, Class.Quant.Grav.
\textbf{12}, 895, (1995).


\end{thebibliography}
\end{document}